\documentclass[journal]{IEEEtran}
\IEEEoverridecommandlockouts
\usepackage{cite}
\usepackage{amsmath,amssymb,amsfonts}
\usepackage{algorithmic}
\usepackage{graphicx}
\usepackage{textcomp}
\usepackage{xcolor}
\usepackage{subfigure}
\usepackage{booktabs}
\usepackage{epstopdf}
\usepackage{footnote}
\usepackage{multicol}
\usepackage{multirow}

\newcommand{\yh}[1]{\textcolor{black}{{#1}}}

\usepackage{geometry}
\usepackage[T1]{fontenc}
\usepackage{aecompl}
\def\BibTeX{{\rm B\kern-.05em{\sc i\kern-.025em b}\kern-.08em
		T\kern-.1667em\lower.7ex\hbox{E}\kern-.125emX}}
\begin{document}

\title{Can Far-field Beam Training Be Deployed for Cross-field Beam Alignment in Terahertz UM-MIMO Communications?\\
}
\author{
    Yuhang~Chen, \IEEEmembership{Student Member,~IEEE}, Chong Han, \IEEEmembership{Senior~Member,~IEEE}, and Emil Bj{\"o}rnson,~\IEEEmembership{Fellow,~IEEE}
    \thanks{    This paper was presented in part at IEEE GLOBECOM, Dec. 2023~\cite{ref_TPBE_GC}.
    
    Y. Chen is with the Terahertz Wireless Communications (TWC) Laboratory, Shanghai Jiao Tong University, Shanghai 200240,
    China (e-mail: yuhang.chen@sjtu.edu.cn).

    C. Han is with Terahertz Wireless Communications (TWC) Laboratory,
    and with the Department of Electronic Engineering and Cooperative Medianet
    Innovation Center (CMIC), Shanghai Jiao Tong University, Shanghai 200240,
    China (e-mail: chong.han@sjtu.edu.cn).

    E. Bj{\"o}rnson is with the Department of Computer Science, KTH Royal Institute of Technology, Stockholm, Sweden (email: emilbjo@kth.se).
		}
	}
	\maketitle
	\thispagestyle{empty}

\begin{abstract}
    Ultra-massive multiple-input multiple-output (UM-MIMO) is the enabler of Terahertz (THz) communications in next-generation wireless networks.
    In THz UM-MIMO systems, a new paradigm of \textit{cross-field} communications spanning from \textit{near-field} to \textit{far-field} is emerging, since the near-field range expands with higher frequencies and larger array apertures.
    Precise beam alignment in cross-field is critical but challenging.
    Specifically, unlike far-field beams that rely only on the angle domain, the incorporation of dual-domain (angle and distance) training significantly increases overhead.
    A natural question arises of whether far-field beam training can be deployed for cross-field beam alignment.
    In this paper, this question is answered, 
    by demonstrating that the far-field training enables sufficient signal-to-noise ratio (SNR) in both far- and near-field scenarios, while exciting all channel dimensions.
    Based on that, we propose a subarray-coordinated hierarchical (SCH) training with greatly reduced overhead.
    To further obtain high-precision beam designs, we propose a two-phase angle and distance beam estimator (TPBE). 
    Extensive simulations demonstrate the effectiveness of the proposed methods.
    Compared to near-field exhaustive search, the SCH possesses 0.2\% training overhead.
    The TPBE achieves 0.01~degrees and 0.02~m estimation root-mean-squared errors for angle and distance.
    Furthermore, with the estimated beam directions, a near-optimal SNR with 0.11~dB deviation is attained after beam alignment. 
    
\end{abstract}

\section{Introduction}
\label{sec_introduction}

The Terahertz (THz) bands, ranging from 0.1 THz to 10 THz, could provide an abundance of bandwidth and are promising for enabling wireless links with extreme capacity for six-generation (6G) communications and beyond~\cite{ref_TSR_THz,ref_6G_Mag}.
At THz frequencies, there exists noticeable signal attenuation, caused by spreading, molecular absorption, and diffusely scattering losses~\cite{ref_hybrid_beamforming}.
Fortunately, the sub-millimeter wavelength allows the incorporation of hundreds and even thousands of antenna elements into one antenna array, leading to ultra-massive multiple-input multiple-output (UM-MIMO) systems.
These systems could assist in establishing highly directional links that efficiently combat the propagation losses in the THz band. 
Furthermore, with beam steering, THz UM-MIMO systems can reach Tbps data rates and enable massive connectivity~\cite{ref_THz_Old_revisit}.

As a combined effect of having a tiny wavelength and a large array aperture (compared to the wavelength), THz UM-MIMO systems introduce a novel \textit{cross-field} communication paradigm where the communication distance spans from the array's radiative \textit{near-field} to \textit{far-field}~\cite{ref_cross,ref_hybrid_THz_CE,ref_6G_near_focusing}.
Strict beam alignment is crucial in cross-field communications, which is the premise of enabling high beamforming gain to compensate for the THz band path losses and guarantee reliable transmissions~\cite{ref_BM_3GPP, ref_THz_Old_revisit, ref_AoSA_training}. 
To find the optimal beam alignment direction, procedures referred to \textit{beam training} and \textit{beam estimation} are usually conducted~\cite{ref_BM_3GPP, ref_AoSA_training}.
Particularly, beam training refers to a process that involves scanning multiple high-gain beams in various directions.
This is useful in mitigating propagation losses encountered when there is no channel state information, thereby facilitating reliable control signal reception. 
After the beam training, beam estimation is executed.
This involves the deployment of different estimation algorithms that determine the most favorable direction for optimal alignment~\cite{ref_Milli_TWC}.

\yh{Beam alignment in THz cross-field communications is challenging, due to the intrinsic high training overhead and difficulties in developing beam estimation.
First, 
unlike far-field beams relying solely on angle-domain resolution, cross-field beams feature angle- and distance-domain resolutions, which inspires the study of dual-domain beam training~\cite{ref_nearfield_book_emil,ref_near_training_UCA}.
However, in UM-MIMO systems, the quantity of high-gain beams in the angle domain scales with the massive number of antennas~\cite{ref_BM_3GPP}. Incorporating distance-domain beams further amplifies the overall beam count for training.
Moreover, in THz UM-MIMO systems, hybrid-beamforming (HBF) architectures designed for hardware efficiency are widely deployed, which contain a restricted number of RF chains,  e.g., $N_{\rm RF} = 16$ RF chains versus $N=1024$ antennas~\cite{ref_hybrid_beamforming, ref_HBF_emil}.
Given the maximum number of beams generated simultaneously by the HBF constrained by the number of RF chains, beam training must be conducted sequentially, thereby enlarging the training overhead.}
\yh{Second, most of the existing beam estimation algorithms focus on far-field scenarios by estimating the alignment angles~\cite{ref_Milli_TWC,ref_power_angle_est,ref_fast_training}, which are inapplicable in cross-field communications where distance estimation is also critical.
Therefore, it is imperative to develop cross-field beam training and estimation with low training overhead and efficiency.}

\subsection{Related Work}

\yh{The beam pattern emerges from the production of the codeword and array response vector in the channel, which is highly dependent on the channel model. 
Therefore, existing literature on beam alignment was designed separately focusing on far-field and near-field scenarios, respectively. 
Within a fixed channel model, beam alignment encompasses three main steps, namely, codebook design, the training process to determine the order of beam searching and the algorithms for beam estimation. 
This is because each beam is generated by a codeword in a predefined codebook, representing one realization of the beamformer. While beam patterns and beam estimation algorithms should be adjusted to channel models.}

\subsubsection{Far-field Beam Alignment}
In far-field scenarios, beam alignment methods are designed based on the planar-wave channel model (PWM).
The phases of the array response vector within the PWM exhibit a linear relationship with spatial angles. 
Consequently, codebooks are designed to generate beams steering at various angles.
One common realization for the angle domain codebook is the spatial Discrete Fourier Transform (DFT) codebook, within which, the beams direct uniformly to discrete grids in the angle domain~\cite{ref_BM_3GPP}, with the beam number equal to the number of antennas. 
However, during the training process, an exhaustive search of beams in the DFT codebook is conducted, which brings high training overhead.
To address this, hierarchical codebooks and training strategies were developed. 
The far-field hierarchical codebook contains beams with different widths in the angle domain.
During the searching process, the beamwidth is progressively narrowed down.
Hierarchical codebooks were studied in various systems, such as uniform planar array (ULA)~\cite{ref_BMW_SS}, array-of-subarrays (AoSA)~\cite{ref_AoSA_training} and quadruple-uniform planar array (QUPA)~\cite{ref_QUPA}.

Beam estimation in far-field could be categorized into power-based and estimation algorithm-based methods.
In the power-based method, alignment direction is determined by the beam pair in the codebook yielding the maximum received power~\cite{IEEE_Std_802_15_3c,ref_BMW_SS, ref_AoSA_training,ref_QUPA}.
Since the beam direction is determined by the directions of beams in the codebook, the power-based estimation usually possesses lower accuracy. 
In contrast, estimation algorithm-based methods involve further signal processing on the received signal with various estimation algorithms, which usually achieve higher accuracy.
For example, in~\cite{ref_fast_training}, compressive sensing (CS)-based estimation was proposed.
In addition, authors in~\cite{ref_Root_MUSIC_HDAPA} and~\cite{ref_Milli_TWC} proposed multiple signal classification (MUSIC) algorithms for AoSA and dynamic AoSA (DAoSA) HBF structures.

\subsubsection{Near-field Beam Alignment}
In near-field scenarios, beam training methods are designed based on the spherical-wave channel model (SWM).
Since phases of the array response vector are related to both the spatial angle and the propagation distance in the SWM~\cite{ref_nearfield_book_emil}, the training codebooks were designed by considering both parameters.
For example, the exhaustive search beam training codebooks were designed for uniform-circular arrays (UCA) systems in~\cite{ref_near_training_UCA} and intelligent reflecting surface (IRS)-assisted 
systems in~\cite{ref_codebookdesign_Dai}.
To reduce the training overhead of the exhaustive search, hierarchical near-field codebooks related to different widths in the angular and distance domains were studied.
For example, two low-overhead hierarchical beam training schemes for near-field MIMO systems were proposed in~\cite{ref_eltraining_near}.
The hierarchical codebook was studied in~\cite{ref_NF_Alexandropoulos} and~\cite{ref_IRS_NF_codebook}.
In addition to the hierarchical codebook, a two-phase beam training method was proposed in~\cite{ref_fast_near_training}, which sequentially determines the angles and distances of the users to reduce the training overhead.

Similar to the far-field, beam estimation in the near-field consists of power- and estimation algorithm-based methods.
In the power-based method with limited accuracy, beam directions were determined by the directions of the beam in the codebook~\cite{ref_eltraining_near,ref_NF_Alexandropoulos,ref_fast_near_training}. 
By contrast, the algorithm-based method could achieve high accuracy.
For example, by exploiting symmetry and statistics of the antenna array, a joint angle and distance estimation method was proposed in~\cite{ref_near_paraest}.
In~\cite{ref_CE_nearfield_Dai} and~\cite{ref_NF_sparse_CE}, the near-field CS-based methods were proposed by designing the near-field channel representation.
In~\cite{ref_hybrid_THz_CE}, the channel parameters are obtained via a fixed point network. 

Since the SWM is the ground truth channel model, beam alignment designed based on SWM can be applied in the cross far- and near-field.
However, compared to far-field training and estimation, the additional distance domain undoubtedly increases the training overhead and computation complexity, despite the codebook, training strategy, and algorithm designs.
This raises the question of whether far-field beam training can be effectively applied to cross-field beam alignment scenarios.

\subsection{\yh{Contributions and Paper Structure}}
Inspired by the limitations of conventional far- and near-field beam alignment approaches, in this work, we propose a far-field training with beam estimation framework for THz cross-field beam alignment.
The proposed framework significantly reduces training overhead through the utilization of far-field training beams, while the employment of precise beam estimation enables near-optimal beam alignment.
Specifically, we first establish the feasibility of employing far-field beams for training.
Beam training aims at enhancing the signal-to-noise ratio (SNR) to guarantee successful control signal reception when no channel state information is available.
By employing the far-field beams, the SNR is sufficiently high in both far- and near-field scenarios due to the limited propagation distances in the near-field and the high array gains in the far-field. 
After that, we propose a subarray-coordinated hierarchical (SCH) beam training method to further reduce the far-field training overhead.
Moreover, we propose a two-phase beam estimator (TPBE), which decouples the estimation of angles and distances in the cross-field with computational efficiency.

In our prior and shorter version of this work~\cite{ref_TPBE_GC}, we focused on a narrow-band system and proposed a corresponding estimation algorithm utilizing a far-field exhaustive search for beam training. 
In this work, we extend the system and algorithms by including wideband features of THz channels and propose the SCH beam training method to reduce the training overhead.
The main contributions of this work are summarized as follows.
\begin{itemize}
    \item \textbf{We propose the far-field training with beam estimation framework and SCH beam training method.}
    We consider a wideband hybrid spherical- and planar-wave channel model (HSPM) and show that far-field training beams ensure adequate SNR for control signal detection in both far- and near-field. This is due to the limited propagation distances in the near-field and high array gains in the far-field. 
    To further reduce the training overhead, we design a two-dimensional (2D) hierarchical codebook for the subarray and propose an SCH training method.
    The SCH training accounts for the beam width and the proximity of alignment angles across different subarrays to decrease the number of training beams required.
    
    \item \textbf{We propose the TPBE with signal reconstruction and estimation algorithm phases for efficient cross-field beam estimation}.
    In the signal reconstruction phase, the TPBE initially separates the angle estimation for different subarrays and distances, which reduces the computational complexity. 
    Then, it employs multi-carrier signal processing, consolidating signals from multiple carriers to enhance the accuracy of estimations compared to single-carrier approaches.
    Regarding the estimation algorithm, TPBE utilizes both the MUSIC and maximum-likelihood estimation (MLE) for determining angles. 
    While MUSIC-based techniques offer superior accuracy at higher transmit power levels, MLE is preferred for its computational efficiency.
    Based on the estimated angles, the phase variation across subcarriers is utilized to obtain the distance.

    \item \textbf{We conduct extensive simulations and performance evaluations.} 
    The metrics of interest include the training overhead, the estimation accuracy in terms of root-mean-squared error (RMSE) for angles and distances, and SNR after alignment, as the primary goal of beam alignment is to maximize SNR.
    Results show that by comparing to the near-field exhaustive search approach, the proposed SCH incurs only 0.2\% of the training overhead.
    The proposed TPBE achieves high precision in angle and distance estimation, reaching RMSE values of $0.01^\circ$ and $0.02$~m, respectively.
    In addition, the SNR after alignment is near-optimal, which outperforms existing algorithms and deviates by only 0.11~dB from the upper bound, i.e., alignment with the real channel, when the transmit power of the pilot signal exceeds 20~dBm.
\end{itemize}

The remainder of the paper is organized as follows.
 The system and channel models are introduced in Sec.~\ref{sec_system_overview}.
 The beam alignment problem, feasibility analysis for deploying far-field beams for training, the  SCH training, and the received signal model are presented in Sec.~\ref{sec_Beam_Alignment_and_Training}.
 The TPBE for cross-field beam estimation is introduced in Sec.~\ref{sec_cross_feild_beam_estimation}.
 Performance evaluation and numerical analysis are provided in Sec.~\ref{sec_Performance_Evaluation}. Finally, the paper is concluded in Sec.~\ref{sec_conclusion}.
 
\textbf{Notation}: 
$a$ is a scalar.
$\mathbf{a}$ denotes a vector. 
$\mathbf{A}$ represents a matrix. 
$\mathbf{I}_{N}$ refers to an identity matrix of dimension $N$.
$\mathbf{A}(i,j)$ depicts the element at the $i^{\rm th}$ row and $j^{\rm th}$ column of $\mathbf{A}$. 
$\mathbf{a}(i)$ denotes the $i^{\rm th}$ element $\mathbf{a}$. 
$(\cdot)^{\mathrm{T}}$ defines the transpose. 
$(\cdot)^{\mathrm{H}}$ refers to the conjugate transpose.
$\mathbb{C}^{M\times N}$ depicts the set of $M\times N$-dimensional complex-valued matrices. 
$\vert\cdot\vert$ calculates absolute values.
${\rm blkdiag}\left\{\cdot \right\}$ denotes block diagonal.
$\circ$ denotes Hadamard product. 
$ {\rm tr}\left\{\cdot \right\}$ calculates trace of a matrix. 

\section{System Overview}
\label{sec_system_overview}

In this section, we first introduce the system model and the HSPM channel model~\cite{ref_HSPM} for THz UM-MIMO cross-field communications.
Then, we demonstrate the applicability of the HSPM to the cross-field. 

\subsection{System Model}
\begin{figure}[t]
    \centering
    {\includegraphics[width= 0.4\textwidth]{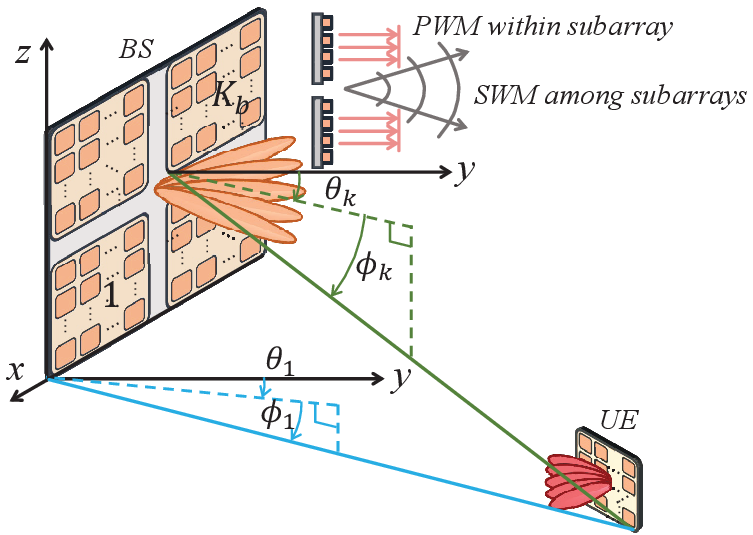}}
    \caption{Generalized THz UM-MIMO system model.}
    \label{fig_system_model}
\end{figure} 

As illustrated in Fig.~\ref{fig_system_model}, we consider a general THz UM-MIMO system model that can be used to describe systems with compact arrays, AoSA, and widely-spaced multiple-subarray (WSMS) HBF structures~\cite{ref_WSMS}. 
The base station (BS) is equipped with a planar UM-MIMO array, which can be configured in either square or rectangular shapes. In this work, we specifically consider a square-shaped array, comprising $N_b$ antennas uniformly divided into $K_b$ subarrays. 
Each subarray contains $N_{ab} = N_b/K_b$ antennas.
Within a subarray, the antenna spacing is $d = \lambda/2$, where $\lambda$ denotes the center carrier's wavelength.
The subarray spacing can be $d$ or multiple of it to represent a general UM-MIMO configuration.
The user equipment (UE) is equipped with a uniform planar array (UPA) containing  $N_u$ antennas. The antenna spacing of the antenna array at UE is fixed at $d$. 

We consider a wideband multi-carrier transmission and an uplink beam training process.
To facilitate the training process, a single pilot stream is considered for each subcarrier~\cite{ref_fast_training}. 
The received signal $y[m]$ at the $m^{\rm th}$ subcarrier on the BS can be represented as
\begin{equation}\label{equ_received_signal_BS_training}
y[m]={\mathbf{w}}^{\mathrm{H}}[m]
\mathbf{H}[m]{\mathbf{f}}[m]{s}+{\mathbf{w}}^{\mathrm{H}}\mathbf{n}[m],~m = 1,\dots,M,
\end{equation}
where $M$ denotes the total number of subcarriers.
The combining vector at the BS is represented as ${\mathbf{w}}[m]=\mathbf{W}_{\rm RF}\mathbf{w}_{\rm BB}[m]\in\mathbb{C}^{N_b}$,
with $\mathbf{W}_{\rm RF}\in\mathbb{C}^{N_b\times N_{\rm RF}}$ denoting the analog combiner, and $N_{\rm RF}$ representing the number of RF chains.
We assume each subarray is connected to one RF chain, i.e., $N_{\rm RF} = K_b$.
Therefore, $\mathbf{W}_{\rm RF}$ takes the block diagonal form
\begin{equation}
    \label{equ_Wrf}
	\mathbf{W}_{\rm RF}=
    {\rm blkdiag}\left\{ \tilde{\mathbf{w}}_{1}, \dots, \tilde{\mathbf{w}}_{K_b}\right\}, 
\end{equation}
where $\tilde{\mathbf{w}}_{k} \in\mathbb{C}^{N_{ab}}$ stands for the combining vector of the $k^{\rm{th}}$ RF chain, for $k = 1,\dots, K_b$. 
The analog combining is implemented by phase shifters, in which each element in $\tilde{\mathbf{w}}_k$ satisfies the constant modulus constraint as $\vert \tilde{\mathbf{w}}_{k}(i)\vert  = \frac{1}{\sqrt{N_{ab}}}$.
The digital combiner for the $m^{\rm{th}}$ subcarrier is depicted as $\mathbf{w}_{\rm BB}[m] \in\mathbb{C}^{N_{\rm RF}}$, while the channel matrix of the $m^{\rm th}$ subcarrier is represented as $\mathbf{H}[m]\in\mathbb{C}^{N_b\times N_u}$. 

At the UE, a fully connected (FC) HBF structure with $N_{\rm RF}$ RF chains is deployed~\cite{ref_hybrid_beamforming}. The beamforming vector ${\mathbf{f}}[m] = \mathbf{F}_{\rm RF}\mathbf{f}_{\rm BB}[m]\in\mathbb{C}^{N_u}$, where $\mathbf{F}_{\rm RF}\in\mathbb{C}^{N_u\times N_{\rm RF}}$ and $\mathbf{f}_{\rm BB}[m] \in\mathbb{C}^{N_{\rm RF}}$
denote the analog and digital beamformers, respectively.
In the FC structure, each RF chain connects to all antennas, which leads that $\mathbf{F}_{\rm RF}$ does not contain zeros as $\mathbf{W}_{\rm RF}$ in~\eqref{equ_Wrf}. 
Each element of $\mathbf{F}_{\rm RF}$ satisfies the constant modulus constraint, $\vert \mathbf{F}_{\rm RF}(i,j)\vert = \frac{1}{\sqrt{N_u}} $.
In addition, the system satisfies the power constraint, i.e., $\Vert \mathbf{F}_{\rm RF}\mathbf{f}_{\rm BB}[m]\Vert^2 = 1$ and $\Vert \mathbf{W}_{\rm RF}\mathbf{w}_{\rm BB}[m]\Vert^2 = 1$.
In~\eqref{equ_received_signal_BS_training}, the pilot signal ${s}$ satisfies $\vert s\vert^2 = P$, where $P$ denotes the transmit power.
Moreover, $\mathbf{n}[m]\in\mathbb{C}^{N_u}$ represents the complex Gaussian noise, following the distribution $\mathcal{N}(0,\sigma^2)$, with $\sigma^2$ denoting the noise variance.


\subsection{Channel Model}
\label{subsec_Channel_Model}
We deploy the HSPM in this work, which employs PWM within subarray and SWM among subarrays~\cite{ref_HSPM}.
It has been demonstrated in~\cite{ref_HSPM} that the HSPM strikes a good balance between high modeling accuracy and low complexity when compared to PWM and SWM, which makes it particularly well-suited for cross-field communications.
In the system depicted in Fig.~\ref{fig_system_model}, the channel matrix for the $m^{\rm th}$ subcarrier $\mathbf{H}[m]$ using HSPM is represented as
\begin{equation}\label{equ_HSPM}
   \begin{split}
    \mathbf{H}[m]  =&\sum_{p=1}^{N_p} \alpha_p[m]\!
    \Big[\mathbf{a}_{N_a} (\psi_{bpx}^{1}, \psi_{bpz}^{1}) [\mathbf{a}_{N_u} (\psi_{upx}^{1}, \psi_{upz}^{1})]^{\rm H}; \\
        \dots;&e^{-j\Delta \Phi_{K_b p}}
            \mathbf{a}_{N_a}(\psi_{bpx}^{K_b},\psi_{bpz}^{K_b})[\mathbf{a}_{N_u}(\psi_{upx}^{K_b},\psi_{upz}^{K_b}]^{\rm H}\Big]\!,
   \end{split}
\end{equation}
where $N_p$ denotes the total number of propagation paths, and $\alpha_p[m]$ represents the complex path gain for the $p^{\rm th}$ path and the $m^{\rm th}$ subcarrier, with $p = 1,\dots, N_p$.
There is $\alpha_p[m] = h_p e^{-j 2\pi m\Delta f \tau_p}$, where $h_p$ denotes the complex coefficient, $\tau_p = 
\frac{r_p}{c}$ represents the delay of the $p^{\rm th}$ path. $r_p$ and $c$ represent the propagation distance and light speed, respectively.

The array response vectors at the BS and UE are denoted as $\mathbf{a}_{N_a}(\cdot,\cdot)\in\mathbb{C}^{N_{ab}}$ and $\mathbf{a}_{N_u}(\cdot,\cdot)\in\mathbb{C}^{N_{u}}$, respectively, which follow a unified form.
To simplify the notation, we consider an $N$ element square-shaped UPA in the $xz$-plane with the azimuth and elevation angle pair $(\theta,\phi)$.
In this case, the array response vector $\mathbf{a}_N(\psi_{x},\psi_{z}) \in\mathbb{C}^{N}$ can be expressed as 
\begin{equation}
    \mathbf{a}_N(\psi_{x},\psi_{z}) = \mathbf{a}_{\sqrt{N}}(\psi_{x}) \otimes \mathbf{a}_{\sqrt{N}}(\psi_{z}), 
\end{equation}
where $\psi_x = \sin\theta\cos\phi $, and $\psi_z = \sin\phi$ represent the spatial angles in the $x$- and $z$-axis, respectively.
Moreover, $\mathbf{a}_{\sqrt{N}}(\psi_{x})$ and $\mathbf{a}_{\sqrt{N}}(\psi_{z})$ both adhere to a unified form as
\begin{equation}
    \label{equ_array_steering_vector}
\mathbf{a}_{\sqrt{N}}(\psi)=
\left[1, \dots, \mathrm{e}^{-j\frac{2\pi }{\lambda}(\sqrt{N} - 1)\psi}\right]^{\mathrm{T}}.
\end{equation}

In~\eqref{equ_HSPM}, subscripts $b$ and $u$ are used to differentiate between the BS and UE, respectively. The superscript $k=1,\dots, K_b$ identifies subarrays at the BS. 
For propagation path $p$, the spatial angles along $x$- and $z$-axis relating to the $k^{\rm th}$ subarray at the BS and UE are represented as $\psi_{bpx}^k = \sin \theta_{bp}^k\cos \phi_{bp}^k $, $\psi_{bpz}^k = \sin \phi_{bp}^k$,  $\psi_{upx}^k=\sin \theta_{up}^k\cos \phi_{up}^k$ and $\psi_{upz}^k = \sin \phi_{up}^k$, respectively, with $(\theta_{bp}^k,\phi_{bp}^k)$ and $(\theta_{up}^k,\phi_{up}^k)$
denoting the physical azimuth and elevation angles pairs at the BS and UE, respectively. 
Moreover, $\Delta \Phi_{kp} = \frac{2\pi}{\lambda}(D^{k}_p-D_p)$ denotes the phase shift term caused by spherical-wave propagation among subarrays.
For the $p^{\rm th}$ path, $D^{k}_p$ represents the distance between UE and the $k^{\rm th}$ subarray at the BS, while $D_p$ depicts the distance between the UE and the reference position locating at the center of the antenna array at the BS, respectively.

The channel matrix $\mathbf{H}[m]$ in~\eqref{equ_HSPM} can be arranged in a more compact form as
\begin{subequations}\label{equ_HSPM_compact}
\begin{align}
    \mathbf{H}[m] &= \sum_{p=1}^{N_p} \alpha_p[m] \mathbf{A}_{bp}\mathbf{A}_{up}^{\rm H},\\
    &= \tilde{\mathbf{A}}_{b} \boldsymbol{\Lambda}[m]\tilde{\mathbf{A}}_{u}^{\rm H},
\end{align}
\end{subequations}
where $\mathbf{A}_{bp} = {\rm blkdiag} \left\{\mathbf{a}_{bp}^{1},\dots,\mathbf{a}_{bp}^{K_b}e^{-j\Delta\Phi_{K_bp}} \right\}\in\mathbb{C}^{N_b\times K_b}$ and $\mathbf{A}_{up} = \left[\mathbf{a}_{up}^{1},\dots, \mathbf{a}_{up}^{K_b} \right]\in\mathbb{C}^{N_u\times K_b}$ denote the array response matrices for the $p^{\rm th}$ path.
Notably, we have denoted $\mathbf{a}_{N_a}(\psi_{bpx}^{k}, \psi_{bpz}^{k})$ as $\mathbf{a}_{bp}^{k}$ and 
$\mathbf{a}_{u}(\psi_{upx}^{k}, \psi_{upz}^{k})$
as $\mathbf{a}_{up}^{k}$ for simplicity. 
The array response matrix at the BS $\tilde{\mathbf{A}}_{b} \in\mathbb{C}^{N_b\times K_bN_p}$, 
where $\tilde{\mathbf{A}}_{b} = {\rm blkdiag}$ $\left\{ \tilde{\mathbf{A}}_{b1},\dots, \tilde{\mathbf{A}}_{bK_b} \right\}$ and $\tilde{\mathbf{A}}_{bk} =  [\mathbf{a}_{b}^{k},\dots,\mathbf{a}^{k}_{bN_p}]\in\mathbb{C}^{N_{ab}\times K_b}$.
The matrix containing the complex path gains is $\boldsymbol{\Lambda}[m] \in \mathbb{C}^{K_bN_p \times K_bN_p} $ with 
$\boldsymbol{\Lambda}[m] = {\rm diag} \{ [\alpha_1[m],\dots, \alpha_{N_p}[m]\dots, \alpha_1[m],\dots, $ $\alpha_{N_p}[m]]\}$.
In addition, $\tilde{\mathbf{A}}_{u}\in\mathbb{C}^{N_u\times K_bN_p}$ is constructed as $\tilde{\mathbf{A}}_{u}=\left[\tilde{\mathbf{A}}_{u1},\dots,\tilde{\mathbf{A}}_{uK_b}\right]$ with $\tilde{\mathbf{A}}_{uk}=\left[\mathbf{a}_{u1}^{k},\dots, \mathbf{a}_{uN_p}^{k} \right] \in\mathbb{C}^{N_u\times N_p}$.

\subsubsection{Angles and Phase Shift}
\label{subsubsec_Angles_Phase_Shift}
For propagation path $p$, the azimuth and elevation angles pair $(\theta_{bp}^k,\phi_{bp}^k)$ in $\mathbf{a}^{k}_{bp} $, $(\theta_{up}^k,\phi_{up}^k)$ in $\mathbf{a}^{k}_{up}$ and the phase shift $\Delta \Phi_{k p}$ in~\eqref{equ_HSPM} vary with the subarray index $k$. This variation arises due to the consideration of the spherical-wave propagation among subarrays in the HSPM.
However, there exist geometric relationships between parameters for the reference and the subsequent subarrays~\cite{ref_HSPM}.
Specifically, for the line-of-sight (LoS) path, we can eliminate the subscript $p$ and superscript $k$ to obtain relationships as
\begin{subequations}
    \label{equ_angles}
    \begin{align}
    \theta_{b}^{k}&\!= \!{\rm arccos}\!{\left(\!\!\frac{ D\!\cos\theta_b\cos\phi_b  }{\sqrt{\!(D\!\cos\phi_b )^2\!+\!\Gamma_{kx}^2\!+\!2D\Gamma_{kx}\sin\theta_b \!\cos\phi_b }}\!\!\right)}\!,
    \\
    \phi_{b}^{k}&\!= \!{\rm arccos}\!{\left(\!\!\frac{ D\!\cos\theta_u  \cos\phi_b}{\sqrt{(D\!\cos\theta_b )^2\!+\!\Gamma_{kz}^2\!+\!2D\Gamma_{kz}\cos\theta_b \!\sin\phi_b }}\!\!\right)}\!,
    \end{align}
\end{subequations}
where $D$ denotes the distance between the reference positions, 
$\Gamma_{kx}$ and $\Gamma_{kz}$ represent the spacings of subarray $k$ relative to the reference position on the $x$- and $z$-axis, respectively.
Moreover, for the LoS path, we have $\theta_{u}^{k} = -\theta_{b}^{k}$ and $\phi_{u}^{k} = -\phi_{b}^{k}$.
Furthermore, the phase shift term $\Delta\Phi_k$ can be calculated as
\begin{equation}
    \label{equ_phase_difference_initial}
    \Delta \Phi_k\!=\!\frac{2\pi D}{\lambda} \!\!\left(\!\! \sqrt{1 \!+\! \frac{2 \!\left( \Gamma_{kz}\varpi_z \! + \!\Gamma_{kx}\varpi_x   \right)  }{D} \! +\!\frac{\Gamma_{kx}^2 \!+\! \Gamma_{kz}^2}{D^2}} \!-\! 1\!\!\right)\!,
\end{equation}
where $\varpi_x  = \sin\theta_b \cos\phi_b $ and $\varpi_z  = \sin\phi_b $ represent the spatial angles at the reference position.
Similar geometric relationships exist for parameters associated with non-line-of-sight (NLoS) paths. In these cases, the distance $D$ in~\eqref{equ_angles} and~\eqref{equ_phase_difference_initial} should be replaced with the distance from the last scatter to the reference position at the antenna array at the BS or UE.

We can observe from~\eqref{equ_angles} and~\eqref{equ_phase_difference_initial} that the parameters in the HSPM, namely, $(\theta_{b}^{k}, \phi_{b}^{k})$ and $\Delta\Phi_k$ are dependent on both the angles $(\theta_{b}, \phi_{b})$ and distance $D$.
In contrast, when utilizing far-field PWM, the angles for different subarrays in~\eqref{equ_angles} are assumed to be identical, denoted as $\theta_b^k=\theta_b$ and $\phi_b^k=\phi_b$.
Furthermore, for PWM, a first-order Taylor expansion is employed to approximate~\eqref{equ_phase_difference_initial}, resulting in the phase shift for PWM denoted as $\Delta \Phi_{k{\text{PWM}}} = \frac{2\pi }{\lambda} \left( \Gamma_{kz}\varpi_z - \Gamma_{kx} \varpi_x\right)$.
Therefore, the parameters based on a planar-wave assumption are only reliant on the angles associated with the reference position.

\subsubsection{Applicable Range}
To demonstrate the applicability range of the HSPM, we analyze the phase differences and amplitude errors between it and the ground-truth SWM.
The channel element between the $l^{\rm th}$ transmitted and $i^{\rm th}$ received antenna for the SWM $\mathbf{H}_{\rm S} \in \mathbb{C}^{N_b\times N_u}$ is expressed as
\begin{equation}\label{equ_SWM_channel}
\mathbf{H}_{\rm S}(i,l)=\Sigma_{p=1}^{N_p}\alpha_p e^{-j\frac{2\pi}{\lambda}(D^{il}_p - D_p)},
\end{equation} 
where, for simplicity, we omit the subcarrier index $m$. 
For the $p^{\rm th}$ path, the communication distance from the $l^{\rm th}$ transmitted antenna to the $i^{\rm th}$ received antenna in~\eqref{equ_SWM_channel} is denoted as $D^{il}_p$, with $i = 1,\ldots, N_t$ and $l = 1,\ldots, N_r$.
Moreover, $D_p$ represents the distance between the reference positions at the BS and UE. 
To simplify the analysis, we remove the subscript $p$ and consider the dominant LoS path in the THz channel. The phase shift term $\frac{2\pi}{\lambda}(D^{il}- D^{11})$ in~\eqref{equ_SWM_channel} can be calculated as
\begin{subequations} \label{equ_phase_difference}
    \begin{align}
     &\frac{2\pi}{\lambda}(D^{il} - D^{11}) \notag\\
    = &\frac{2\pi}{\lambda}
    \bigg\{ D^{11}
    \Big[ 1+\frac{2(d_{i_x} + d_{l_x})\varpi_x + 2(d_{i_z} + d_{l_z})d\varpi_z}{D^{11}} \notag\\
    &+ \frac{(d_{i_x} + d_{l_x})^2 + (d_{i_z} + d_{l_z})^2}{(D^{11})^2}\Big]^{\frac{1}{2}} -D^{11}\bigg\},\\
    \approx&\frac{2\pi}{\lambda}\bigg[ (d_{i_x} + d_{l_x})\varpi_x + (d_{i_z} + d_{l_z})d\varpi_z  \notag\\
    & +\frac{(d_{i_x} + d_{l_x})^2 + (d_{i_z} + d_{l_z})^2}{2D^{11}} \\
    &+ \frac{[(d_{i_x} + d_{l_x})\varpi_x+ (d_{i_z} + d_{l_z})\varpi_z]^2}{2D^{11}}\bigg]\notag
    ,
    \end{align}
    \end{subequations}
where for the $i^{\rm th}$ antenna at the BS, $d_{i_x}, d_{i_z}$ 
represent the distances of the antenna to the reference position at the center of the antenna array on the $x$- and $z$-axis, respectively. For the $l^{\rm th}$ antenna at the UE, $d_{l_x}$ and $d_{l_z}$ denote the distances to the reference position at the center of the antenna array on the $x$- and $z$-axis, respectively.
The result is obtained by taking the second-order Taylor expansion.

Given the fact that the HSPM considers spherical-wave propagation among subarrays, the phase difference for the channel element between the reference positions of a subarray at the BS and UE is zero.
Therefore, the phase difference for the HSPM is comparable to that of a single subarray at the UE and BS, leading to the maximum values of $d_{i_x}$ and $d_{i_z}$ as $d_{i_x}=d_{i_z}= \frac{S_{ab}}{2}$.
Here, $S_{ab} = \frac{\lambda}{2}\sqrt{2N_{ab}}$ denotes the aperture of the subarray at the BS.
While the maximum values of $d_{l_x}$ and $d_{l_z}$ are $d_{l_x}=d_{l_z}=\frac{S_{u}}{2}$, where $S_{u} = \frac{\lambda}{2}\sqrt{2N_{u}}$ represents the aperture at the UE.
By setting $\varpi_x= \varpi_z=0$, we can maximize the phase difference. 
Using a common value of $\pi/8$ to determine when phase variations can be neglected~\cite{ref_Fraunhofer}, we obtain $D^{11} =D_{Fa}= \frac{2(S_{ab} + S_u)^2}{\lambda}$, 
where $D_{Fa}$ denotes the Fraunhofer distance of a subarray.
Therefore, when $D$ is larger than $D_{Fa}$, the phase error in the HSPM becomes negligible.

The received signal amplitude is inversely proportional to the communication distance.
According to~\cite{ref_nearfield_book_emil}, we can calculate the relative difference between the edge and center of the array, to determine that the amplitude difference can be neglected when $D > 2\sqrt{S_b^2 + S_u^2}=D_B$, where $S_b$ denotes the array aperture of the entire array at the BS, and $D_B$ represents the Bj{\"o}rnson distance.
Therefore, we can conclude that the HSPM applies to the range with $D >\max[D_{Fa}, D_B]$.
It is worth noticing that the PWM approximates the wave propagation as plane for the entire array.
The number of parameters to describe the PWM is proportional to $N_p$.
However, by considering the phase errors, the PWM is only applicable to a distance greater than the Fraunhofer distance of the antenna array, i.e., $D_{F}= \frac{2(S_{b} + S_u)^2}{\lambda}$.
If we also consider the amplitude difference, we can conclude that the PWM applies within the range where $D_{11} >\max[D_{F}, D_B]$.

For example, we consider the system illustrated in Fig.~\ref{fig_system_model}, operating at a central frequency $f=0.3$~THz.
At the BS, a THz WSMS UM-MIMO system is employed with $N_{\rm RF}=K_b=4$ RF chains and subarrays, each comprising $N_{ab}=256$ antennas. 
The subarray spacing is set as $128\lambda$.
The FC structure at the UE contains $N_u=64$ antennas. 
In this case, $D_{Fa}\approx 0.48$~m, $D_B \approx 0.40$~m, $D_{F}\approx 85.85$~m.
Therefore, the HSPM is applicable for a range larger than $0.48$~m. 
In summary, the HSPM is suitable for communication distances ranging from near- to far-field, i.e., the cross-field communication scenario. 
A comparative analysis of different channel models is presented in TABLE~\ref{Tab_applicable_range}.

\begin{table}[t]
	\centering
	\caption{Comparison of different channel models.}
	\begin{tabular}{ccc}
		\toprule
        Model&  Parameters & Applicable range \\
	    \midrule
		SWM &$\propto N_b N_u$ & All distances\\
        HSPM & $\propto K_b N_p$ & $>\max[D_{Fa},D_B]$, e.g., $>0.48$~m\\
        PWM & $\propto N_p$& $>\max[D_{F},D_B]$, e.g., $>85.85$~m\\
		\bottomrule
	\end{tabular}
 \label{Tab_applicable_range}
\end{table}


\section{Beam Alignment and Training}
\label{sec_Beam_Alignment_and_Training}
In this section, we first formulate the beam alignment problem and demonstrate the feasibility of deploying far-field beams for cross-field training.
Next, we introduce a far-field exhaustive search beam training technique by considering the HSPM. To reduce the training overhead associated with that approach, we propose the SCH beam training method for THz cross far- and near-field communications.

\subsection{HSPM Beam Alignment Problem Formulation}
\label{subsec_HSPM_Beam_Alignment_Problem_Formulation}
Beam alignment aims at finding the pair of precoder and combiner to maximize the received power or SNR, as~\cite{ref_BM_3GPP}
\begin{equation}
    \label{equ_training_opt}
    \begin{split}
    &\left\{ \mathbf{w}_{\rm opt}[m], \mathbf{f}_{\rm opt}[m]\right\} = \arg\max \left\vert \mathbf{{w}}^{\rm H}[m] \mathbf{H}[m]\mathbf{{f}}[m]\right \vert^2, \\
    &{\rm s.t.~}  \vert \tilde{\mathbf{w}}_{k}(i)\vert  = \frac{1}{\sqrt{N_{ab}}},\Vert \mathbf{W}_{\rm RF}\mathbf{w}_{\rm BB}[m]\Vert^2 = 1,\\
    &~~~~~ \vert \mathbf{F}_{\rm RF}(i,j) \vert = \frac{1}{\sqrt{N_u}}, \Vert \mathbf{F}_{\rm RF}\mathbf{f}_{\rm BB}[m]\Vert^2 = 1.
    \end{split}
\end{equation}
Since the THz channel exhibits a LoS dominant property, where the power of the LoS path greatly surpasses the cumulative power of other paths~\cite{ref_THz_Old_revisit}, we focus on the alignment of the LoS path and eliminate the subscript $p$ to denote the LoS path.
It is worth noticing that we still consider a multi-path channel while focusing on aligning to the direction of the LoS path.
In this case, by considering free-space path loss, noise power $\sigma^2$, and  isotropic antennas, the SNR can be calculated as
$\frac{P\lambda^2 G_u G_b}{(4\pi  \sigma D)^2},$
which is proportional to the beamforming gains $G_u$ and $G_b$ at the UE and BS while inversely proportional to $D^2$.
To achieve beam alignment, $G_u$ and $G_b$ should be maximized, which is achieved when $\mathbf{w}_{\rm opt}[m]$ and $\mathbf{f}_{\rm opt}[m]$ are align to the LoS path's angle and distance, i.e., $\mathbf{W}_{\rm RF}=\frac{1}{\sqrt{N_{ab}}}
\left[\begin{matrix}
{\mathbf{a}}_{b}^{1}&\dots&\mathbf{0};\dots;
\mathbf{0}&\dots&{\mathbf{a}}_{b}^{K_b}e^{-j\Delta\Phi_{K_b}}
\end{matrix}\right]$,
$\mathbf{F}_{\rm RF}=\frac{1}{\sqrt{N_u}}
\left[\begin{matrix}
{\mathbf{a}}_{u}^{1},\dots, {\mathbf{a}}_{u}^{K_b}
\end{matrix}\right]$,
$\mathbf{w}_{\rm BB}[m] = \mathbf{f}_{\rm BB}[m]=\frac{1}{\sqrt{K_b}}\mathbf{1}_{K_b}$. 

\subsection{Can We Use Far-field Beam Training in the Cross-field?}

\begin{figure}[t]
    \centering
    \subfigure[Beamforming gain.]{\includegraphics[width=0.23\textwidth]{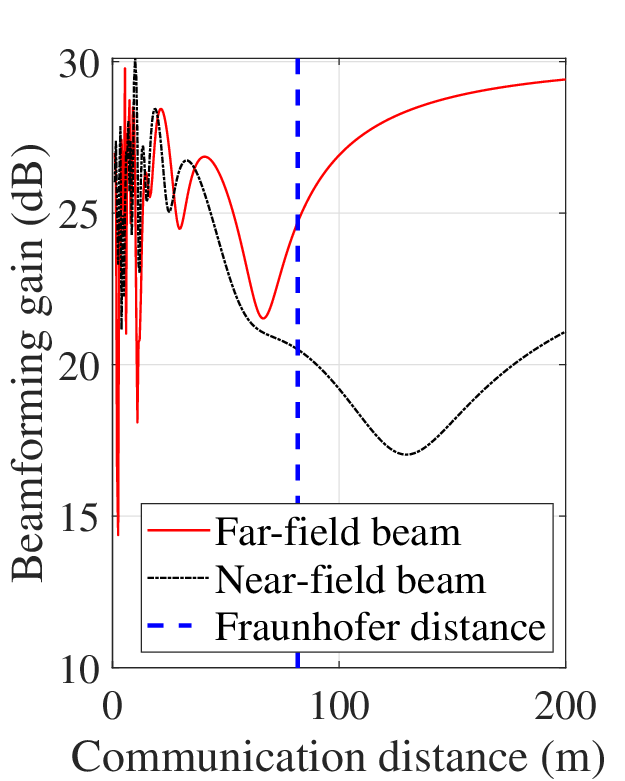} }
    \subfigure[SNR.]{\includegraphics[width=0.23\textwidth]{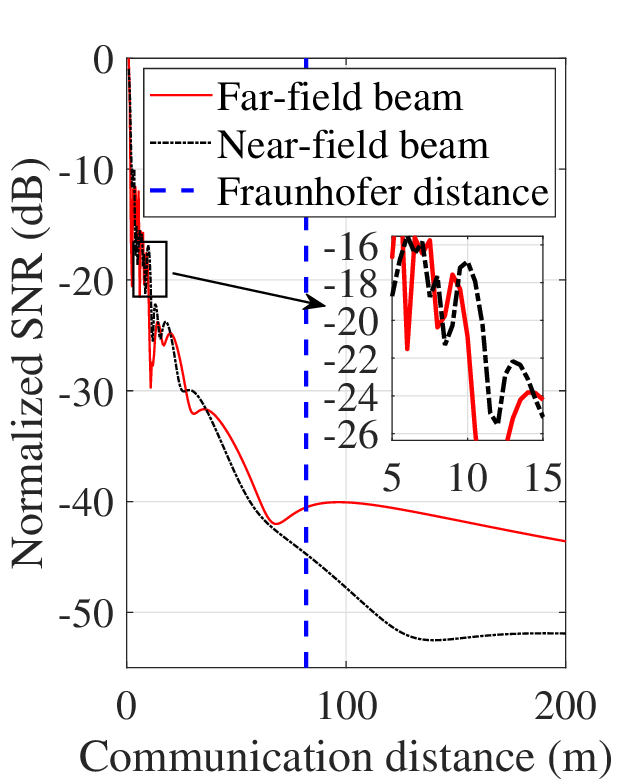} }
    \caption{Beamforming gain and SNR comparison by using far-field and near-field beams.}
    \label{fig_farnear_comp}
\end{figure}

As analyzed in Sec.~\ref{subsec_Channel_Model}, the channel parameters in the HSPM $(\theta_{b}^{k}, \phi_{b}^{k})$ and $\Delta\Phi_k$ in HSPM vary with both angles and distance. 
In contrast, when considering the PWM, these parameters are only dependent on angles.
Therefore, due to beam misalignment, directly training and communicating with far-field beams designed based on the PWM 
can result in reduced beamforming gain and received power.
An example of this is provided in Fig.~\ref{fig_farnear_comp}(a), where a 1024-element WSMS with 4 subarrays and $128\lambda$ subarray spacing are deployed at the BS. The UE with one antenna is positioned at 10~m distance from the BS. By deploying the near-field beam to align to the angle and distance, the beamforming gain reaches the maximum of $1024\approx30$~dB. In contrast, the beamforming gain by deploying the far-field beam that only aligns to the angle is 26~dB, resulting in $50\%$ less received power than that with near-field beams.
To address this, a straightforward approach is to deploy the near-field beam training designed to cover different angles and distances, as commonly found in the literature~\cite{ref_near_training_UCA, ref_codebookdesign_Dai, ref_eltraining_near, ref_NF_Alexandropoulos, ref_IRS_NF_codebook, ref_fast_near_training}.
However, using near-field training introduces additional distance searching, which in turn increases the training overhead.

In this study, we propose to deploy far-field training, and then design supplementary beam estimation techniques that allow us to achieve beam alignment even with far-field training methods.
Since any near-field channel is spanned by a range of far-field channels, it is the beam estimation that must be adapted to the near-field, not the training signaling.
To illustrate the viability of deploying far-field beams for training, we consider the SNR by using far- and near-field beams as depicted in Fig.~\ref{fig_farnear_comp}(b).
Indeed, the far-field beam is imprecise and cannot provide the highest beamforming gain at short distances, i.e., values of $G_u$ and $G_b$ are small, leading to SNR degradation when compared to the near-field beam.
However, by using far-field beams in cases where the communication distance is short, particularly when it falls below the Fraunhofer distance to be in the near-field, e.g., 82~m in Fig.~\ref{fig_farnear_comp}(b), the SNR is sufficiently high by comparing to far-field cases when the distance exceeds the Fraunhofer distance.
In addition, when the communication distance lies in the far-field, the SNR of the far-field beam is larger than that of the near-field beam due to large values of $G_u$ and $G_b$.

Therefore, we conclude that far-field beams can be employed for training, which ensures the SNR for training in both far- and near-field. 
Nonetheless, compared to deploying near-field beams for communications, employing far-field beams still leads to significant SNR losses within the near-field range.  
To address this issue, we propose cross-field beam estimation algorithms in Sec.~\ref{sec_cross_feild_beam_estimation}. 
Based on the estimation results, the BS and UE generate beams to achieve beam alignment.

\subsection{SCH Beam Training}
\label{subsec_SCH_Beam_Training}

\begin{figure*}[t]
    \centering
    \subfigure[Layer 1.]{\includegraphics[width=0.3\textwidth]{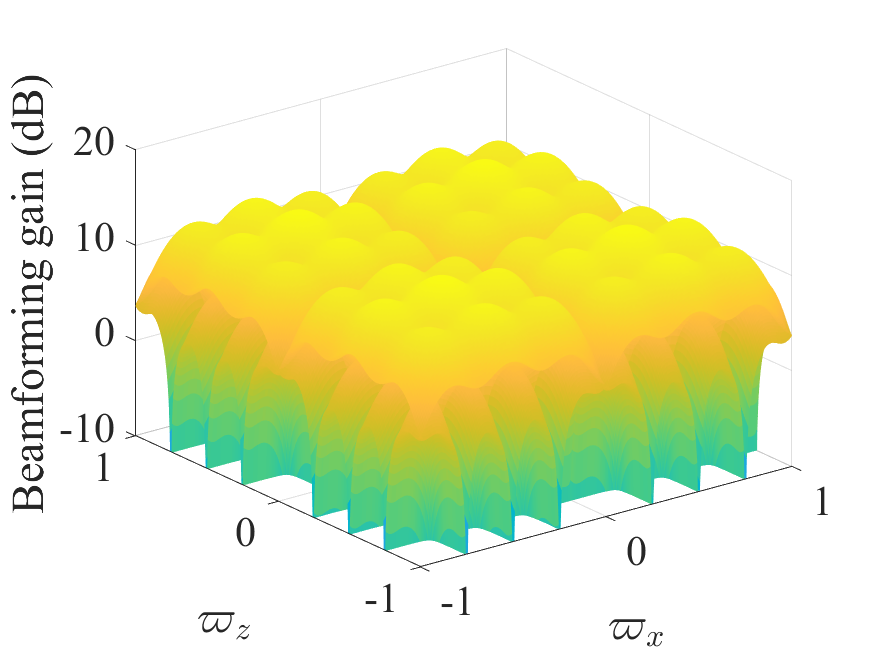} }
    \subfigure[Layer 2.]{\includegraphics[width=0.3\textwidth]{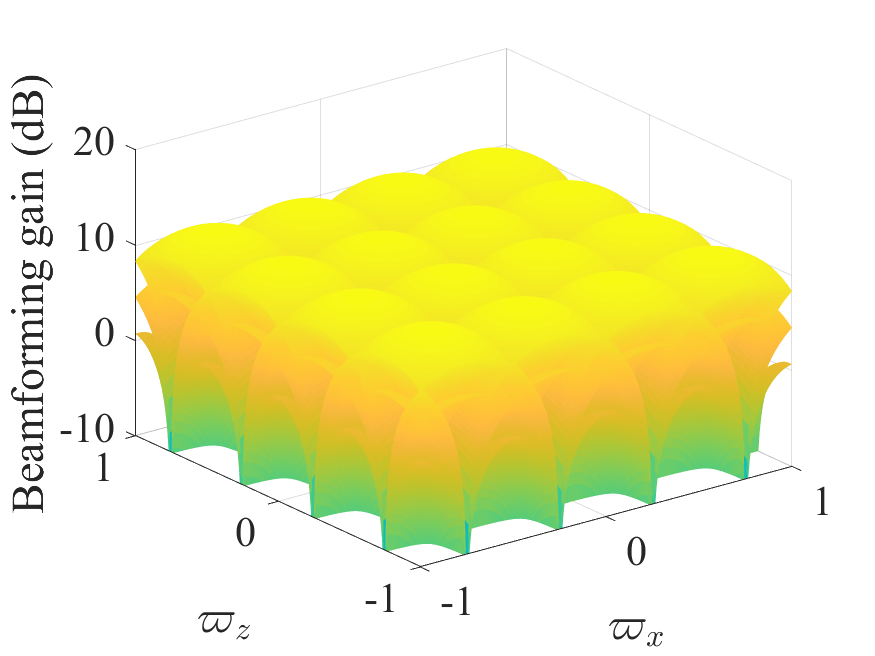} }
    \subfigure[Layer 3.]{\includegraphics[width=0.3\textwidth]{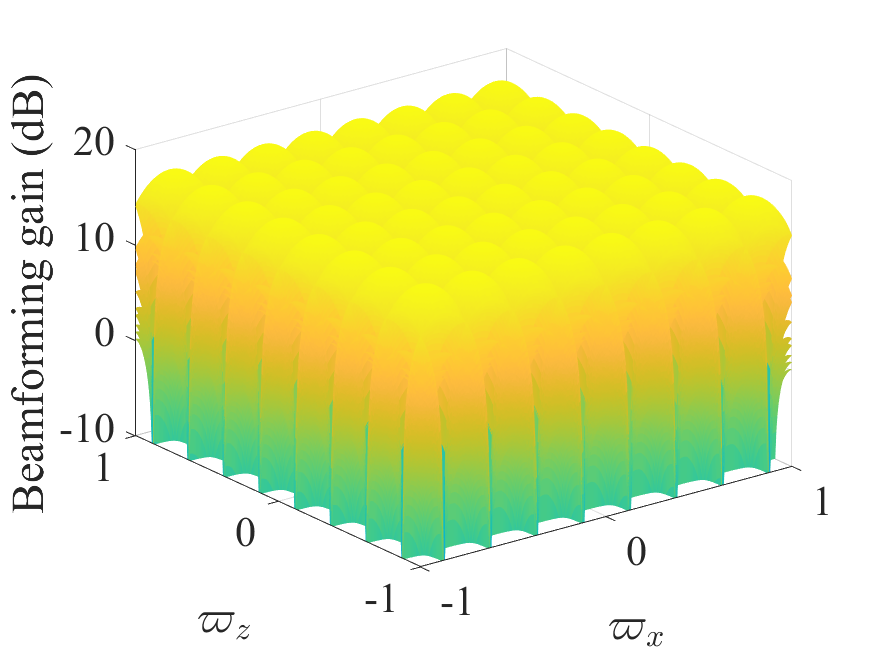} }
    \caption{Beam patterns across various layers of the hierarchical codebook for a 64-element subarray.}
    \label{fig_beampattern}
\end{figure*}

To start with, we demonstrate a far-field exhaustive search-beam training method tailored for HSPM. 
Similar to the exhaustive search with PWM~\cite{ref_BM_3GPP}, the exhaustive search based on the HSPM relies on a DFT codebook that uniformly partitions the spatial angles into $N_b$ grids, with each grid representing a discrete beam direction.
The codebook construction should adapt to the form of HSPM in~\eqref{equ_HSPM}. 
At the BS, the exhaustive search codebook is formed as
\begin{equation}
    \label{equ_W_exh}
    \mathcal{W}_{\textrm{exh}} =\left\{ \mathbf{w}_{1}, \dots, \mathbf{w}_{N_{b}}\right\},
\end{equation}
where $\mathbf{w}_{n} = \frac{1}{\sqrt{K_{b}}}{\rm blkdiag} \Big\{\mathbf{a}_{N_a}\Big(\frac{2(n_x-1)}{N_x}-1,\frac{2(n_z-1)}{N_z}-1 \Big),\dots,$ $\mathbf{a}_{N_a}\Big( \frac{2(n_x-1)}{N_x}-1,\frac{2(n_z-1)}{N_z}-1 \Big)e^{-j\Delta\tilde{\Phi}_{K_b}} \Big\} \mathbf{1}_{K_b} $, $n=(n_x-1)N_x+N_z$, with $n_x = 1,\dots, N_x$, $n_z = 1,\dots, N_z$, $N_x = N_z = \sqrt{N_b}$.
The phase shift term is denoted as
$\Delta \tilde{\Phi}_k = \frac{2\pi }{\lambda}  \left[ \Gamma_{kz}\left(\frac{2(n_z-1)}{N_z}-1\right) - \Gamma_{kx} \left(\frac{2(n_x-1)}{N_x}-1\right) \right]$.
The DFT codebook at the UE can be constructed similarly. 
Notably, $\mathcal{W}_{\textrm{exh}}$ contains $N_b$ codewords. Due to the massive number of antennas in the THz UM-MIMO, the required time to scan all beams in the codebook increases linearly with the number of antennas, which brings high training overhead. 

To address this problem, we propose the SCH training approach described below. 
The SCH first separates the training process for each subarray, with each subarray conducting a hierarchical search. 
Then, the training beams for different subarrays are chosen based on spatial correlations of beams in the hierarchical codebook. 
Specifically, we first separate beam training on different subarrays, each subarray conducts exhaustive search-based training with the DFT codebook, obtained as
\begin{equation}
\begin{split}
    \label{ref_W_suba}
    \mathcal{W}_{\textrm{suba}} &=  \left\{ \mathbf{w}_1, \dots, \mathbf{w}_{N_a}\right\},
\end{split}
\end{equation}
where $\mathbf{w}_{n} = \Xi_{N_a}(:,n)$.
The $N$ element DFT matrix 
\begin{equation}
    \label{equ_DFT_matrix}
\begin{split}
    \Xi_{N} &= \Bigg[\mathbf{a}_{N}(-1, -1),\dots, \\
    &\mathbf{a}_{N}\Big( \frac{2(n_x-1)}{N_x}-1,\frac{2(n_z-1)}{N_z}-1 \Big),\dots, \\
    &\mathbf{a}_{N}\left( \frac{2(N_x-1)}{N_x}-1, \frac{2(N_z-1)}{N_z}-1 \right)\Bigg],
\end{split}
\end{equation}
with $N = N_zN_z$. 
In this subarray-wise exhaustive search, each subarray searches $N_a$ beams, and $K_b$ subarrays perform beam searching sequentially. The total training overhead is obtained as $N_b = K_b N_a$.
Existing studies have proposed many hierarchical training methods to reduce the training overhead.
In this work, we extend the hierarchical codebook in~\cite{ref_BMW_SS} suitable for ULA to two dimensions and adopt it to the planar-shaped subarrays in the WSMS.

The codewords with different widths for each subarray are constructed as follows.
Specifically, we consider each beam in the previous layer to be divided into 4 beams in the next layer, with 2 beams on the $x$- and $z$-axis, respectively. 
Therefore, for subarray $k$, the codebook contains a maximum of $L = \log_4N_a$ layers.
For the $l^{\rm th}$ layer, $l = 1,\dots,\log_4N_a$, there are $2^l$ beams on $x$-axis and $z$-axis, respectively. 
For the $k^{\rm th}$ subarray, the combining vector in the $l^{\rm th}$ layer can be represented as $\tilde{\mathbf{w}}_k(l,b_x,b_z) = \tilde{\mathbf{w}}_{kx}(l,b_x)\otimes \tilde{\mathbf{w}}_{kz}(l,b_z)$, with $b_x=1,\dots,2^l$ indexes the beam on $x$-axis, $b_z=1,\dots,2^l$ indexes the beam on $z$-axis. 
The codeword on the $x$- and $z$-axis are denoted as $\tilde{\mathbf{w}}_{kx}(l,b_x)$ and $\tilde{\mathbf{w}}_{kz}(l,b_z)$, respectively.
By setting $\mathbf{W}_{\rm RF} = {\rm blkdiag}\left\{ \mathbf{0}_{N_a}, \dots, \tilde{\mathbf{w}}_k(l,b_x,b_z),\dots,\mathbf{0}_{N_a}\right\}$ and $\mathbf{w}_{\rm BB} = \mathbf{1}_{K_b}$, the combining vector for the entire array can be obtained as 
${\mathbf{w}}_k(l,b_x,b_z)= \left[ \mathbf{0}_{N_a}; \dots; \tilde{\mathbf{w}}_k(l,b_x,b_z),\dots;\mathbf{0}_{N_a}\right]\in\mathbb{C}^{N_b}$.

In the last layer with $l = \log_4N_a$, the codebook is directly assembled employing $\mathcal{W}_{\textrm{suba}}$ in~\eqref{ref_W_suba}.
In this case, the beams contained in the codebook exhibit the narrowest width $ \frac {2}{\sqrt{N_a}}$ along both $x$-axis and $z$-axis.
In other layers with $l = \log_4N_a - \tilde{l}$, where $\tilde{l} = 1,\dots, \log_4N_a-1$, we partition the subarray into $Q=4^{\lfloor(\tilde{l}+1)/2 \rfloor}$ virtual subarrays, each axis containing $Q_a=2^{\lfloor(\tilde{l}+1)/2 \rfloor}$ virtual subarrays. 
The number of antennas on each axis of the virtual subarray equals $N_{va}=\sqrt{N_a}/Q_a$.
To effectively control the beamwidth of beams within each layer, we define the number of effective subarrays on each axis as $N_A=Q_a$.
When $\tilde{l}$ is odd, and $N_A=Q_a/2$ when $\tilde{l}$ is even.
The codeword $\tilde{\mathbf{w}}_{kx}(l,b_x)$ and $\tilde{\mathbf{w}}_{kz}(l,b_z)$ are constructed in a manner analogous to that described in~\cite{ref_BMW_SS}.
Finally, we compile the codewords ${\mathbf{w}}_k(l,b_x,b_z)$ to form the hierarchical search codebook represented as
\begin{equation}
    \label{equ_W_hie}
    \mathcal{W}_{\textrm{hie}} \!= \!\left\{ {\mathbf{w}}_k(1,1,1), \dots, {\mathbf{w}}_k\left( \log_4N_a,\sqrt{N_a},\sqrt{N_a}\right) \right\}\!.
\end{equation}

It is noteworthy that the codeword, denoted as $\tilde{\mathbf{w}}_k(l,b_x,b_z)$ exhibits a beamwidth $\frac{Q_a^2}{\sqrt{N_a}}$ when $\tilde{l}$ is odd along $x$- and $z$-axis.
The central angles corresponding to the beams along $x$- and $z$-axis are given by $(b_x-1)\frac{Q_a^2}{\sqrt{N_a}} + \frac{Q_a^2}{2\sqrt{N_a}}$, $(b_z-1)\frac{Q_a^2}{\sqrt{N_a}} + \frac{Q_a^2}{2\sqrt{N_a}}$.
These beams encompass ranges of $\left[ (b_x-1)\frac{Q_a^2}{\sqrt{N_a}},b_x\frac{Q_a^2}{\sqrt{N_a}} \right]$ and $\left[ (b_z-1)\frac{Q_a^2}{\sqrt{N_a}},b_z\frac{Q_a^2}{\sqrt{N_a}} \right]$, respectively.
Similarly, when $\tilde{l}$ is even, the beam widths along both axes are $\frac{2Q_a^2}{\sqrt{N_a}}$.
The central angles on the $x$- and $z$-axis are $(b_x-1)\frac{2Q_a^2}{\sqrt{N_a}} + \frac{Q_a^2}{\sqrt{N_a}}$, $(b_z-1)\frac{2Q_a^2}{\sqrt{N_a}} + \frac{Q_a^2}{\sqrt{N_a}}$.
These beams cover ranges of $\left[ (b_x-1)\frac{2Q_a^2}{\sqrt{N_a}},b_x\frac{2Q_a^2}{\sqrt{N_a}} \right]$ and $\left[ (b_z-1)\frac{2Q_a^2}{\sqrt{N_a}},b_z\frac{2Q_a^2}{\sqrt{N_a}} \right]$.
An example of beam patterns for a 64-element subarray with three codebook layers is presented in Fig.~\ref{fig_beampattern}. 
In the figure, we show the aggregation of all beams in each layer.
Notably, lower troughs can be utilized to differentiate between distinct beams.
It can be observed that as the layer number increases, the beamwidth narrows while the beamforming gain is enhanced.
Moreover, each beam originating from the preceding layer is further subdivided into four narrower beams within the subsequent layer.

Based on the constructed codebook, hierarchical beam training could be conducted from subarray to subarray. 
However, as elaborated in Sec.~\ref{subsubsec_Angles_Phase_Shift}, the parameters of subarrays are dependent on each other.
When azimuth and elevation angles are known for the reference position, it becomes possible to extrapolate a potential range of angles for the remaining subarrays.
This revelation serves as the cornerstone for further reducing the training overhead of the SCH.
In particular, based on the geometric relationships in Fig.~\ref{fig_system_model}, there are 
\begin{subequations}
    \label{equ_potential_range}
    \begin{align}
        \varpi_{kx} &= \frac{D_1\varpi_{1x} + \Gamma_{kx}}{D_k},\\
        \varpi_{kz} &= \frac{D_1\varpi_{1z} + \Gamma_{kz}}{D_k}.
    \end{align}
\end{subequations}
Moreover, we approximate the subsequent relationship
\begin{equation}
    D_k\approx D_1 + \frac{2 \left( \Gamma_{kz}\varpi_z-\Gamma_{kx}\varpi_x  \right)}{D_1}.
\end{equation}

By considering the spatial ranges of beams in the final layer for the reference subarray, e.g., the first subarray, we can calculate the potential ranges of $ \varpi_{kx} $ and $ \varpi_{kx}$, by varying the value of $D_1$ from 1 to infinity. The beam training for other subarrays could start with the beam with the narrowest width to cover the potential range.
Steps for implementing the SCH beam training approach are summarized in Algorithm 1.

\begin{table}[t]
	\centering
	\renewcommand
	\arraystretch{} 
	\begin{tabular}{l}
		\toprule
		\textbf{Algorithm 1:} SCH Beam Training \\
		\midrule 
		\textbf{Input}: Number of antennas at the BS $N_b$, number of subarrays $K_b$\\
        1.~Construct hierarchical codebook $\mathcal{W}_{\textrm{hie}}$ \\
        2.~\textbf{For} $k=1:K_b$\\
        3.~~~~~~\textbf{If} $k=1$\\
        4.~~~~~~~~~~~Perform hierarchical beam training based on $\mathcal{W}_{\textrm{hie}}$\\
        5.~~~~~~~~~~~Record the beam range in the last layer\\
        6.~~~~~~\textbf{else}\\
        7.~~~~~~~~~~~Calculate potential range of $\varpi_{kx}$ and $\varpi_{kz}$ by~\eqref{equ_potential_range}\\
        8.~~~~~~~~~~~Start hierarchical beam training with the beam with \\ 
        ~~~~~~~~~~~~~the narrowest width to cover the potential range\\
        9.~~~~~~\textbf{end}\\
        10.~\textbf{end}\\
		\bottomrule
	\end{tabular}
\end{table}

\subsection{Channel Observation}
After beam training, we proceed to collect the received signals, thereby constructing a channel observation tailored for beam estimation. 
We consider that each codeword ${\mathbf{w}}_k(l,b_x,b_z)$ corresponds to the $b_k^{\rm th}$ beam that utilized for training within the $k^{\rm th}$ subarray, $b_k = 1, \dots, B_k$, with $B_k$ representing the total number of training beams for subarray $k$.
We represent ${\mathbf{w}}_k(l,b_x,b_z)$ and $\tilde{\mathbf{w}}_k(l,b_x,b_z)$ 
from the hierarchical codebook in~\eqref{equ_W_hie} as ${\mathbf{w}}_{b_k}$ and $\tilde{\mathbf{w}}_{b_k}$, respectively. 
Moreover, for each codeword at the BS, the UE employs $N_u$ beams for training, which are determined by a DFT codebook denoted as $\mathcal{F} = \left\{ \mathbf{f}_1, \dots, \mathbf{f}_{N_u}\right\}$, 
with $\mathbf{f}_{n_u} = \Xi_{N_{u}}(:,n_u)$.
The matrix $\Xi_{N_{u}}$ is constructed in accordance with~\eqref{equ_DFT_matrix}.
 In this case, $N_u$ signals as~\eqref{equ_received_signal_BS_training} can be collected to obtain $\mathbf{y}_{b_k}[m]\in\mathbb{C}^{1\times N_u}$ as
\begin{subequations}\label{equ_received_signal_BS}
	\begin{align}
	\mathbf{y}_{b_k}[m]&=\sqrt{P}\mathbf{w}^{\mathrm{H}}_{b_k}
	\mathbf{H}[m]\overline{\mathbf{F}}+\mathbf{w}^{\mathrm{H}}_{b_k}\mathbf{n}[m]\\
    &=\sqrt{P}\tilde{\mathbf{w}}^{\mathrm{H}}_{b_k}
	\mathbf{H}_k[m]\overline{\mathbf{F}}+{\mathbf{w}}^{\mathrm{H}}_{b_k}\mathbf{n}[m],
	\end{align}
\end{subequations}
where we consider the pilot signal $s=\sqrt{P}$.
Moreover, $\mathbf{H}_k[m] = \tilde{\mathbf{A}}_{bk} \boldsymbol{\Lambda}_k[m]\tilde{\mathbf{A}}_{uk}$, and $\boldsymbol{\Lambda}_k[m] = \boldsymbol{\Lambda}((k-1)N_p+1:kN_p,(k-1)N_p+1:kN_p)[m]$, which are based on~\eqref{equ_HSPM_compact}.
Notably,~\eqref{equ_received_signal_BS} distinguishes the received signal across different subarrays, as a direct outcome of the proposed SCH training. 

By aggregating all received signals from the $k^{\rm th}$ subarray, we derive the channel observation $\mathbf{Y}_k[m]\in\mathbb{C}^{B_k\times N_u}$ as 
\begin{equation}\label{equ_observation_subarray}
    \begin{split}
    \mathbf{Y}_k[m]=\sqrt{P}\overline{\mathbf{W}}^{\rm{H}}_k
    \mathbf{H}_k[m]\overline{\mathbf{F}}+\overline{\mathbf{N}}_k[m],
    \end{split}
\end{equation}
where $\overline{\mathbf{W}}^{\rm{H}}_k = \left[ \tilde{\mathbf{w}}_{1},\dots, \tilde{\mathbf{w}}_{B_k}\right] \in\mathbb{C}^{N_a\times B_k}$, and $ \overline{\mathbf{N}}_k[m]= \left[ {\mathbf{w}}^{\mathrm{H}}_{1}\mathbf{n}[m], \dots, {\mathbf{w}}^{\mathrm{H}}_{B_k}\mathbf{n}[m] \right]$.
When all received signals from different subarrays are combined, we obtain the overall channel observation $\mathbf{Y}[m]\in\mathbb{C}^{B\times N_u}$, as expressed by 
\begin{equation}
    \label{equ_observation}
    \mathbf{Y}[m]=\left[ {\mathbf{Y}}_{1}[m];\dots;{\mathbf{Y}}_{K_b}[m] \right],
\end{equation}
where $B =\sum_{k-1}^{K_b}B_k $ denotes the total number of beams deployed at the BS during the beam training.

\section{Cross-field Beam Estimation}
\label{sec_cross_feild_beam_estimation}

Given the channel observation matrix in~\eqref{equ_observation}, our objective is to estimate a set of five unknown vectors characterized by $4K_b+1$ parameters, namely $\left(\boldsymbol{\theta}_b,\boldsymbol{\theta}_u,\boldsymbol{\phi}_b,\boldsymbol{\phi}_u,{D} \right)$, 
where $\boldsymbol{\theta}_{b} = [\theta_{b}^1,\dots,\theta_{b}^{K_b} ]^{\rm T}$ and 
$\boldsymbol{\theta}_{u} = [\theta_{u}^1,\dots,\theta_{u}^{K_b} ]^{\rm T}$ contain azimuth angles at the BS and UE,
$\boldsymbol{\phi}_{b} = [\phi_{b}^1,\dots,\phi_{b}^{K_b} ]^{\rm T}$ and 
$\boldsymbol{\phi}_{u} = [\phi_{u}^1,\dots,\phi_{u}^{K_b} ]^{\rm T}$ consist of elevation angles at the BS and UE, respectively. 
Existing beam estimation methods, such as the MLE~\cite{ref_MLE}, typically involve an exhaustive search within a $(4K_b+1)-$dimensional continuous domain for these parameters, which inherently leads to high computational complexity.
In this section, we propose the TPBE approach, which effectively decomposes the beam estimation task into two distinct sub-problems, i.e., angle estimation and communication distance estimation.
Moreover, we propose multi-carrier signal processing for each sub-problem to improve the estimation accuracy. 
These sub-problems are then addressed sequentially, mitigating the computational burden associated with the direct search method in the literature.

\subsection{Angle Estimation}
\label{subsec_angle_estimation}
We first conduct angle estimation based on~\eqref{equ_observation}. 
However, directly estimating the angles still entails a $4K_b-$dimensional search, which introduces substantial computational overhead.
Therefore, we propose to decouple the angle estimation for different subarrays, leveraging the channel observations from different subarrays as outlined in~\eqref{equ_observation_subarray}.

Particularly, we can vectorize $\mathbf{Y}_k[m]$ to obtain $\check{\mathbf{y}}_{k}[m] = {\rm vec} \left( \mathbf{Y}_k[m]\right) \in \mathbb{C}^{B_k N_u}$, which can be expressed as 
\begin{subequations}
\label{equ_hat_yk}
\begin{align}
    \check{\mathbf{y}}_{k}[m]&=\sqrt{P}\boldsymbol{\Psi}_k^{\rm H}\mathbf{E}_k \boldsymbol{\alpha}[m]+ \check{\mathbf{n}}_k[m],   
\end{align}
\end{subequations}
where $\check{\mathbf{n}}_k[m] ={\rm vec} \left( \overline{\mathbf{N}}_k[m]\right) \in \mathbb{C}^{B_k N_u}$ represents the received noise vector, and $\boldsymbol{\Psi}_k = \overline{\mathbf{W}}_k \otimes
\overline{\mathbf{F}}^{*} \in\mathbb{C}^{N_aN_u\times B_kN_u}$. 
The array response matrix, encompassing the production of array response vectors at both the BS and UE, is defined as $\mathbf{E}_k = \left[\mathbf{a}_{b1}^k\otimes (\mathbf{a}_{u1}^k)^*,\dots, \mathbf{a}_{bN_p}^k\otimes (\mathbf{a}_{uN_p}^k)^*\right] \in \mathbb{C}^{N_aN_u \times N_p}$.
Moreover, $\boldsymbol{\alpha}[m] = \left[{\alpha}_1[m], \dots, {\alpha}_{N_p}[m] \right]^{\rm T} \in\mathbb{C}^{N_p}$. 
For received signals at different subcarrier $m$, we aggregate them together to obtain $ \check{\mathbf{Y}}_{k} = \left[ \check{\mathbf{y}}_{k}[1], \dots, \mathbf{y}_{k}[M]\right] \in\mathbb{C}^{B_k N_u\times M}$, where  
\begin{equation}
    \label{equ_signal_suba_subc}
    \check{\mathbf{Y}}_{k} = \boldsymbol{\Psi}_k^{\rm H}\mathbf{E}_k \mathbf{S}+ \check{\mathbf{N}}_k,
\end{equation}
with $\mathbf{S} = \left[ \boldsymbol{\alpha}[1],\dots, \boldsymbol{\alpha}[M]\right] \in \mathbb{C}^{N_p\times M }$ denoting the equivalent received signal, $\check{\mathbf{N}}_k = \left[ \check{\mathbf{n}}_k[1], \dots,\check{\mathbf{n}}_k[M]  \right]\in \mathbb{C}^{N_p\times M }$ representing the collected noise.

\begin{table}[t]
	\centering
	\renewcommand
	\arraystretch{} 
	\begin{tabular}{l}
		\toprule
		\textbf{Algorithm 2:} TPBE Angle Estimation\\
		\midrule 
		\textbf{Input}: Channel observation matrix $\mathbf{Y}[m]$ in~\eqref{equ_observation}\\
        1~\textbf{For} $k = 1: K_b$ \\
        2.~~~~~~Follow the steps from~\eqref{equ_hat_yk} to~\eqref{equ_signal_suba_subc} to construct $\check{\mathbf{Y}}_{k}$\\
        3.~~~~~~Construct the search spectrum by~\eqref{equ_DoA_MLE} or~\eqref{equ_DoA_MUSIC}\\
        4.~~~~~~Deploy 2D-GSS to obtain $\left(\hat{{\theta}}_b^k,\hat{{\phi}}_b^k\right) $\\
        6.~\textbf{end}\\
        \textbf{Output}: Estimated angles $\left(\hat{\boldsymbol{\theta}}_b,\hat{\boldsymbol{\phi}}_b\right) $\\
		\bottomrule
	\end{tabular}
\end{table}

Based on the signal model in~\eqref{equ_signal_suba_subc}, we first deploy the MLE-based angle estimator as
\begin{equation}
    \label{equ_DoA_MLE}
    \left(\hat{{\theta}}_b^k,\hat{{\phi}}_b^k\right) = \arg\max_{ \theta, \phi} 
    {\rm tr}\left\{ \frac{\left[ \left(\boldsymbol{\Psi}_k^{\rm H} \mathbf{e}(\theta, \phi)\right)^{\rm H}\check{\mathbf{Y}}_{k} \right]^2 }{ \left[\boldsymbol{\Psi}_k^{\rm H} \mathbf{e}(\theta, \phi)\right]^2}\right\},
\end{equation}
where the vector for spectrum searching is constructed as 
$\mathbf{e}(\theta, \phi)=\mathbf{a}_{N_a}(\sin\theta\cos\phi,\sin\phi) \otimes \mathbf{a}^*_{N_u}(-\sin\theta\cos\phi,-\sin\phi)$.
This choice is guided by the fact that $\theta_{u1}^k = -\theta_{b1}^k$ and $\phi_{u1}^k = -\phi_{b1}^k$ for the LoS path.
Moreover, since the LoS path possesses the highest power, searching the spectrum of~\eqref{equ_DoA_MLE} could yield the estimation of ${{\theta}}_b^k$ and ${{\phi}}_b^k$.

The estimation accuracy of the MLE-based estimator in~\eqref{equ_DoA_MLE} is limited at high SNRs, inspiring the exploration of estimators with better performance. 
Therefore, we introduce a MUSIC-based angle estimator as an alternative.
Given the reconstructed observation $\check{\mathbf{Y}}_k$ in~\eqref{equ_signal_suba_subc}, the covariance matrix is calculated as 
\begin{equation}\label{equ_covariance_matrix}
\mathbf{R}_{\mathrm{y}_k}=\check{\mathbf{Y}}_k \check{\mathbf{Y}}_k^{\rm H}.
\end{equation}
Then, we conduct eigenvalue decomposition on $\mathbf{R}_{\mathrm{y}_k}$ to obtain
\begin{equation}\label{equ_EVD}
\mathbf{R}_{\mathrm{y}_k}=
\mathbf{U}_{\mathrm{s}}\boldsymbol{\Sigma}_{\mathrm{s}}\mathbf{U}_{\mathrm{s}}^{\mathrm{H}}+\mathbf{U}_{\mathrm{n}}\mathbf{\Sigma}_{\mathrm{n}}\mathbf{U}_{\mathrm{n}}^{\mathrm{H}},
\end{equation}
where the matrix $\mathbf{U}_{\mathrm{s}}\in\mathbb{C}^{B_kN_u\times N_p}$ contains the $N_p$ eigenvectors of $\mathbf{R}_{\mathrm{y}_k}$ that spans the signal subspace, 
$\boldsymbol{\Sigma}_{\mathrm{s}}\in\mathbb{C}^{N_p\times N_p}$ comprises the $N_p$ leading eigenvalues, and.
Moreover, $\boldsymbol{\Sigma}_{\mathrm{n}}\in\mathbb{C}^{(B_kN_u-N_p)\times (B_kN_u-N_p)}$ contains the remaining eigenvalues, with $\mathbf{U}_{\mathrm{n}}\in\mathbb{C}^{B_kN_u\times (B_kN_u-N_p)}$ containing the corresponding eigenvectors. 
Since the leading eigenvalues are significantly larger than the rest, we can identify $N_p$ by analyzing the eigenvalues.
Based on this, DoA estimation can be obtained using the following MUSIC-based estimator~\cite{ref_Milli_TWC} as
\begin{equation}
    \label{equ_DoA_MUSIC}
    \left(\hat{{\theta}}_b^k,\hat{{\phi}}_b^k\right) = \arg\max_{ \theta, \phi} 
    \left[\mathbf{e}^{\mathrm{H}}(\theta,\phi)\mathbf{U}_{\mathrm{n}}\mathbf{U}_{\mathrm{n}}^{\mathrm{H}}\mathbf{e}(\theta,\phi)\right]^{-1}.
\end{equation}

To efficiently search for the angle estimates by~\eqref{equ_DoA_MLE} and~\eqref{equ_DoA_MUSIC} with reduced complexity, we employ a two-dimensional golden section search (2D-GSS) method, as described in~\cite{ref_Milli_TWC}.
It is noteworthy that the angle search within the 2D-GSS can begin within the angular range determined by the SCH beam training process described in Sec.~\ref{subsec_SCH_Beam_Training}, which further reduces the computational complexity.
Furthermore, angle estimation for different subarrays can be conducted simultaneously, making efficient use of parallel processing capabilities.
The complete procedures for the proposed TPBE angle estimation are summarized in Algorithm 2.

\subsection{Distance Estimation}
\label{subsec_distance_estimation}

\begin{table}[t]
	\centering
	\renewcommand
	\arraystretch{} 
	\begin{tabular}{l}
		\toprule
		\textbf{Algorithm 3:} TPBE Distance Estimation\\
		\midrule 
		\textbf{Input}: Channel observation matrix $\mathbf{Y}[m]$ in~\eqref{equ_observation},\\
        ~~~~~~~~~estimated angles $\left(\hat{\boldsymbol{\theta}}_b,\hat{\boldsymbol{\phi}}_b\right)$ by \textbf{Algorithm 2}\\
        1.~Construct $\check{\mathbf{Y}}$ as~\eqref{equ_check_Y}\\
        2.~Construct the search spectrum by~\eqref{equ_dist_est}\\
        3.~Deploy GSS to obtain $\hat{\tau}$, calculate $\hat{D}=c\hat{\tau}$\\
        \textbf{Output}: Estimated distance $\hat{D}$\\
		\bottomrule
	\end{tabular}
\end{table}

Once we have obtained the angle estimates, it becomes feasible to calculate the communication distance, denoted as $D$, based on the relationship defined in~\eqref{equ_phase_difference_initial}.
Nevertheless, it is crucial to consider that even slight estimation errors in $\theta_b^k$ and $\phi_b^k$ are likely to diverge in the estimation of $D$.
In the sequel, we introduce a robust distance estimation algorithm that leverages the phase shift among different subcarriers to mitigate the impact of estimation errors and ensure reliable distance estimation.

Specifically, the signal models from different subarrays, as that in~\eqref{equ_signal_suba_subc}, can be combined to form a single observation matrix as $\check{\mathbf{Y}} = \left[ \check{\mathbf{Y}}_{1}; \dots; \check{\mathbf{Y}}_{K_b}\right] \in\mathbb{C}^{BN_u\times M}$, where
\begin{equation}
    \label{equ_check_Y}
    \check{\mathbf{Y}} = \boldsymbol{\Psi}^{\rm H}\mathbf{E}\mathbf{S} + \check{\mathbf{N}}, 
\end{equation}
with matrices $\boldsymbol{\Psi} = {\rm blkdiag} \left\{ \boldsymbol{\Psi}_1,\dots,\boldsymbol{\Psi}_{K_b}\right\} \in\mathbb{C}^{N_bN_u \times BN_u} $, $\mathbf{E} = \left[ \mathbf{E}_1;\dots;\mathbf{E}_{K_b}\right]\in\mathbb{C}^{N_bN_u \times N_p}$ and $\check{\mathbf{N}} = \left[ \check{\mathbf{N}}_1;\dots; \check{\mathbf{N}}_{K_b}\right] \in\mathbb{C}^{BN_u \times M}$. 
Furthermore, we can express $\mathbf{S}$ defined in~\eqref{equ_signal_suba_subc} as
\begin{subequations}
\begin{align}
    \mathbf{S} 
    & = \left[\begin{matrix}
        \alpha_{1}^*[1] \mathbf{b}_{1}^{\rm T} ,\dots,
        \alpha_{N_p}^*[1]\mathbf{b}_{{N_p}}^{\rm T} 
    \end{matrix}\right]^{\rm H}\\
    & = \boldsymbol{\Lambda}^*[1]\mathbf{B}^{\rm H},
\end{align}
\end{subequations}
where $\mathbf{b}_{p} = \left[1,\dots,e^{j2\pi (M-1)\Delta f \tau_{p}} \right]^{\rm T} \in\mathbb{C}^{M}$, $\alpha_{p}[1] = h_1 e^{-j 2\pi \Delta f \tau_1}$.
In addition, $\mathbf{B} = \left[\mathbf{b}_{1},\dots, \mathbf{b}_{N_p} \right]\in\mathbb{C}^{N_p\times M}$.
Based on this, we could further reconstruct $\check{\mathbf{Y}}^{\rm H}$ as 
\begin{subequations}
    \label{equ_checkY_H}
\begin{align}
    \check{\mathbf{Y}}^{\rm H} &= \mathbf{B}\boldsymbol{\Lambda}^{*}[1]\mathbf{E}^{\rm H} \boldsymbol{\Psi} + \check{\mathbf{N}}^{\rm H} \\
    &=\sum_{p=1}^{N_p}\alpha_{p}[1]\mathbf{b}_{p}\mathbf{e}_{p}^{\rm H} \boldsymbol{\Psi} + \check{\mathbf{N}}^{\rm H},
\end{align}
\end{subequations}
where $\mathbf{e}_{p}= \left[ \mathbf{a}^1_{bp}\otimes (\mathbf{a}_{up}^{1})^*;\dots; \mathbf{a}^{K_b}_{bp}\otimes (\mathbf{a}_{up}^{K_b})^*\right] \in\mathbb{C}^{N_bN_u}$. 

In~\eqref{equ_checkY_H}, it is observed that the distance information for the LoS path is embedded $\tau_1$ in $\mathbf{b}_{1}$. To estimate $\tau_1$, consequently, $D_1$, we propose constructing a matrix with a similar structure as~\eqref{equ_checkY_H} and performing spectrum searching. 
We begin by generating an estimated received signal $\hat{\mathbf{y}}_{k}\in\mathbb{C}^{B_kN_u}$ based on the estimated angles  $(\hat{\boldsymbol{\theta}}_b,\hat{\boldsymbol{\phi}}_b) $ as 
\begin{equation}
    \label{equ_signal_suba_subc_est}
    \hat{\mathbf{y}}_{k} = \boldsymbol{\Psi}_k^{\rm H}
    \mathbf{e}(\hat{{\theta}}_b^k,\hat{{\phi}}_b^k).
\end{equation}
Then, we collect $\hat{\mathbf{y}}_{k}$ for different subarrays to obtain $\hat{\mathbf{y}} = \left[\hat{\mathbf{y}}_{1};\dots; \hat{\mathbf{y}}_{K_b}\right]\in\mathbb{C}^{BN_u}$.
The estimation of $\tau_1$ or equivalently, $D_1$ can be determined by solving the following problem
\begin{equation}
    \label{equ_dist_est}
    \begin{split}
    &\hat{\tau}_1 = \arg\max_{\tau} {\rm tr} \left( \check{\mathbf{Y}}^{\rm H} \hat{\mathbf{y}}^{\rm H}\mathbf{b}(\tau) \right),
    \end{split}
\end{equation}
where $\mathbf{b}(\tau) = \left[1,\dots,e^{j2\pi (M-1)\Delta f \tau} \right]^{\rm T} \in\mathbb{C}^{M}$.
By examining various values of $\tau$ in the region $[0,\frac{1}{\Delta f})$, we can obtain the estimation result of~\eqref{equ_dist_est}.
We directly conduct a one-dimensional search to the problem in~\eqref{equ_dist_est}.
Finally, the distance estimation $\hat{D}$ is obtained as $\hat{D} = c\hat{\tau}$, where $c$ is the speed of light. The procedures for TPBE angle and distance estimation are summarized in Algorithm~3. Finally, beam alignment can be accomplished based on the estimated angles and distances.

\begin{table}[t]
	\centering
	\caption{Simulation parameters.}
	\begin{tabular}{ccc}
		\toprule
        \textbf{Notation}& \textbf{Definition} & \textbf{Value}\\
		\midrule
		$f_c$ &Carrier frequency& 0.3 THz\\
        $\Delta f$ & Subcarrier spacing &3840 kHz\\
        $M$ & Number of subcarriers & 64, 128\\
        $N_b$ & Number antennas at the BS & 256, 64\\
        $K_b$ & Number subarrays at the BS & 4\\
        $N_{\rm RF}$ & Number RF-chains at the BS and UE & 4\\
        $N_u$ & Number antennas at the UE & 64, 16\\
        $N_p$ & Number of multi-path in the channel & 8\\
		\bottomrule
	\end{tabular}
 \label{Tab_simulation_para}
\end{table}

The time complexity of the MLE-based angle estimator is calculated as $\mathcal{O}\left(B_kN_{a}N_u^2M\right)$, while the time complexity of the MUSIC-based angle estimator is calculated as $\mathcal{O}\left(B_k^2N_u^2M\right)$.
These complexities are significantly reduced in comparison to the complexity associated with a direct search implementation of MLE, obtained as $\mathcal{O}\left(B^2N_u^2M^2\right)$~\cite{ref_MLE}.

\section{Performance Evaluation}
\label{sec_Performance_Evaluation}
In this section, we evaluate the performance of the proposed beam alignment scheme with existing solutions in the literature.

\subsection{Simulation Setup}
We consider the system in Fig.~\ref{fig_system_model}, where the BS is equipped with the THz WSMS UM-MIMO system, with the subarray spacing fixed at $128\lambda$. The UE is equipped with an FC HBF structure. 
The main simulation parameters are listed in TABLE~\ref{Tab_simulation_para}.
The subcarrier spacing is selected based on the physical layer numerology for beyond 52.6~GHz communications in~\cite{ref_PHY_526G}.
In terms of the channel model, the HSPM in~\eqref{equ_HSPM} is considered with the existence of the LoS path.
The complex path gains, azimuth and elevation angles and communication distances are randomly generated based on the statistical QuaDriGa channel generator~\cite{ref_QuaDRiGa}, which are adjusted by the results measured from an outdoor atrium scenario in the THz band~\cite{ref_Atrium}.
All results are obtained by averaging over 5000 trials of Monte Carlo simulations.

\subsection{Training Overhead}

\begin{table}[t]
	\centering
	\caption{Training overhead comparison.}
	\begin{tabular}{ccc}
		\toprule
        \textbf{Method}& \textbf{Training overhead}  &\textbf{Value}\\
		\midrule
		Near-field exhaustive &$N_uN_b  C_d$ & $4.1\!\times \!10^5$\\
        Far-field exhaustive & $N_uN_b $ & $4.1\!\times \!10^3$\\
        Far-field subarray-wise exhaustive &$N_uN_b $ & $4.1\!\times \!10^3$\\
        Proposed SCH & $4N_uK_b\log_4N_{ab} \!- \!C$ & $768\!-\!C$\\
		\bottomrule
	\end{tabular}
 \label{Tab_training_overhead}
\end{table}

We begin by comparing the training overheads of different methods.
The results are summarized in Table~\ref{Tab_training_overhead}, where $C_d=100$ indicates the number of beams in the distance domain for each angle. 
Values in the table are computed based on $N_b = 256$, $K_b = 4$ and $N_u = 16$.
Moreover, $C$ is an integer within $[0, 4(K_b-1)\log_4 N_a] = [0, 36]$, whose value varies with angles in the channel. 
The results demonstrate that the proposed SCH methodology incurs a significantly reduced training overhead. \yh{Specifically, it requires only 0.2\% and 18.7\% of the overhead generated by the near-field and far-field exhaustive search methods, respectively.}

\subsection{Estimation Accuracy}

To evaluate the estimation accuracy, we consider the estimation RMSE for angles and distances. The RMSE for angle estimation in degree is calculated as
\begin{equation}
    \label{equ_RMSE_A}
\begin{split}
{\textrm{RMSE}}(A)&\!=\!\frac{1}{2K_b}\left\{ \!\!\sqrt{\mathbb{E}\left[\sum_{k=1}^{K_b}(\hat{{\theta}}_{b}^{k}-{\theta}_{b}^{k})^{2}\!+\!(\hat{{\phi}}_{b}^{k}-{\phi}_b^{k})^{2}\!\right]}\right.
\\
&+\!\!\left.\sqrt{\mathbb{E}\left[\sum_{k=1}^{K_b}(\hat{{\theta}}_{u}^{k}-{\theta}_{u}^{k})^{2}\!+\!(\hat{{\phi}}_{u}^{k}-{\phi}_u^{k})^{2}\right]}\right\}.
\end{split}
\end{equation}
The RMSE for distance estimation in meters is defined as
\begin{equation}
    \label{equ_RMSE_D}
{\textrm{RMSE}}(D)=\sqrt{\mathbb{E}[(\hat{D}-{D})^2]}.
\end{equation}

\begin{figure}[t]
    \centering
    {\includegraphics[width= 0.4\textwidth]{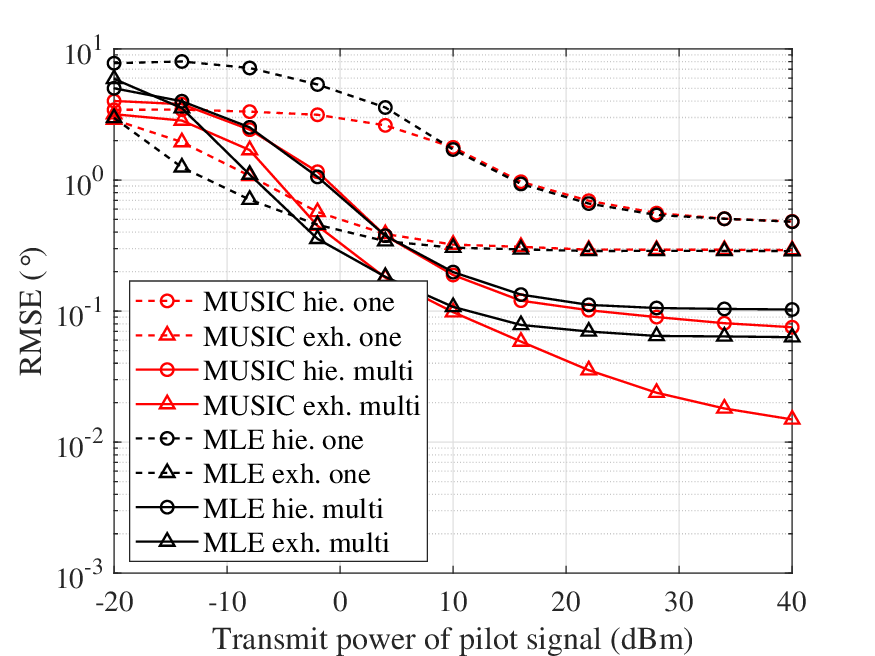}}
    \caption{Angle estimation accuracy with different configurations.}
    \label{fig_ErrCompAngle_OwnAlorithms}
\end{figure}

In Fig.~\ref{fig_ErrCompAngle_OwnAlorithms}, we evaluate the performance of the proposed beam training and beam estimation methods.
The evaluations encompass both subarray-wise exhaustive and SCH hierarchical searches with single and multiple subcarrier settings. 
To facilitate a rigorous comparison, the total transmit power remains invariant for scenarios with single or multiple subcarriers.
Moreover, the number of antennas for each subarray at the BS and the UE is set as 64, and the number of subcarriers is configured as 128.
As illustrated in Fig.~\ref{fig_ErrCompAngle_OwnAlorithms}, 
first, our proposed algorithms with multiple subcarriers achieve lower RMSE values, particularly when the transmit power exceeds $-2$~dBm.
This arises from the treatment of signals across different subcarriers as multiple channel observations, as explicated in~\eqref{equ_observation}. 
Specifically, the MLE algorithm, when employed with a hierarchical search strategy and multiple subcarriers, yields an RMSE that is $0.55^\circ$ lower than for one subcarrier at 22~dBm transmit power.
Second, the exhaustive search strategy generally incurs a lower RMSE than the proposed SCH hierarchical search method. 
This is attributed to the substantial beam count involved in the beam training phase.
However, the magnitude of this RMSE difference is marginal, amounting to a mere $0.04^\circ$ at 40~dBm transmit power for the MLE algorithm with multiple subcarriers.
This negligible performance demonstrates the efficacy of the SCH hierarchical search strategy, which offers reduced training overhead.
Third, by comparing MLE and MUSIC-based algorithms, we conclude that the latter provides higher estimation accuracy, especially at transmit power levels above 4~dBm. \
This is owing to the inherent higher accuracy of the MUSIC-based algorithm.
For example, under conditions of hierarchical search and multiple subcarriers at 16~dBm transmit power, the RMSE for the MUSIC method is approximately halved compared to the MLE approach.

\begin{figure}[t]
    \centering
    {\includegraphics[width= 0.4\textwidth]{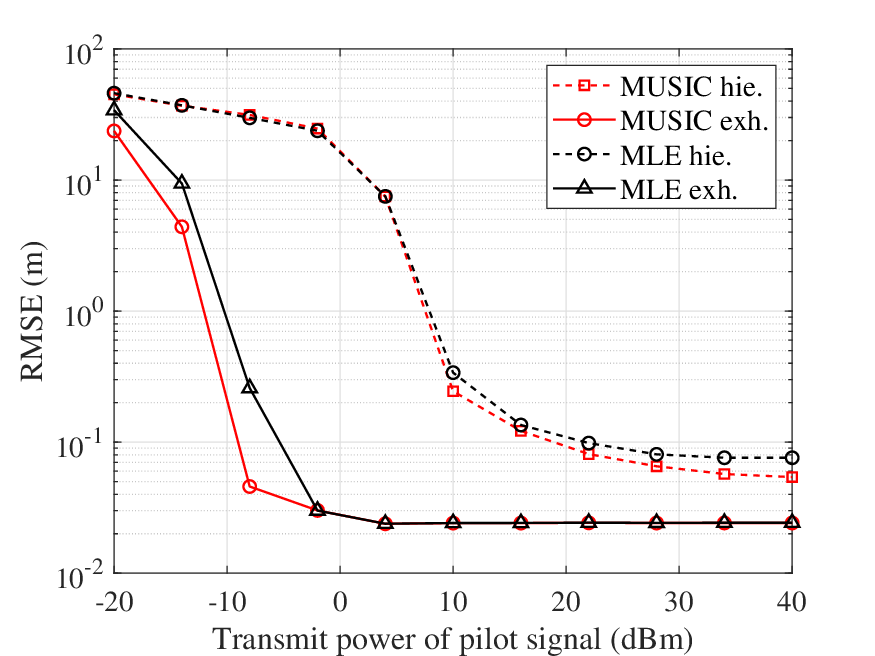}}
    \caption{Distance estimation accuracy with different configurations.}
    \label{fig_ErrCompDist_OwnAlgorithms}
\end{figure}

Next, we compare the distance estimation performance based on the MUSIC and MLE angle estimation results and proposed SCH hierarchical and exhaustive searches, respectively. 
Given that the distance estimation in Sec.~\ref{subsec_distance_estimation} relies on phase differences among multiple subcarriers, our evaluation is confined to scenarios with multiple subcarriers.
As illustrated in Fig.~\ref{fig_ErrCompDist_OwnAlgorithms}, the RMSE for distance estimation is around 
0.08~m when employing hierarchical search, and approximately 0.02~m with the exhaustive search method, for transmit power levels exceeding 20~dBm.
This demonstrates the high accuracy of our proposed methodologies.
Furthermore, the trends observed in distance estimation are similar to those in angle estimation. 
These include an increase in accuracy correlating with higher transmit power, as well as superior performance when employing exhaustive search and MUSIC-based algorithms.

\begin{figure}[t]
    \centering
    {\includegraphics[width= 0.4\textwidth]{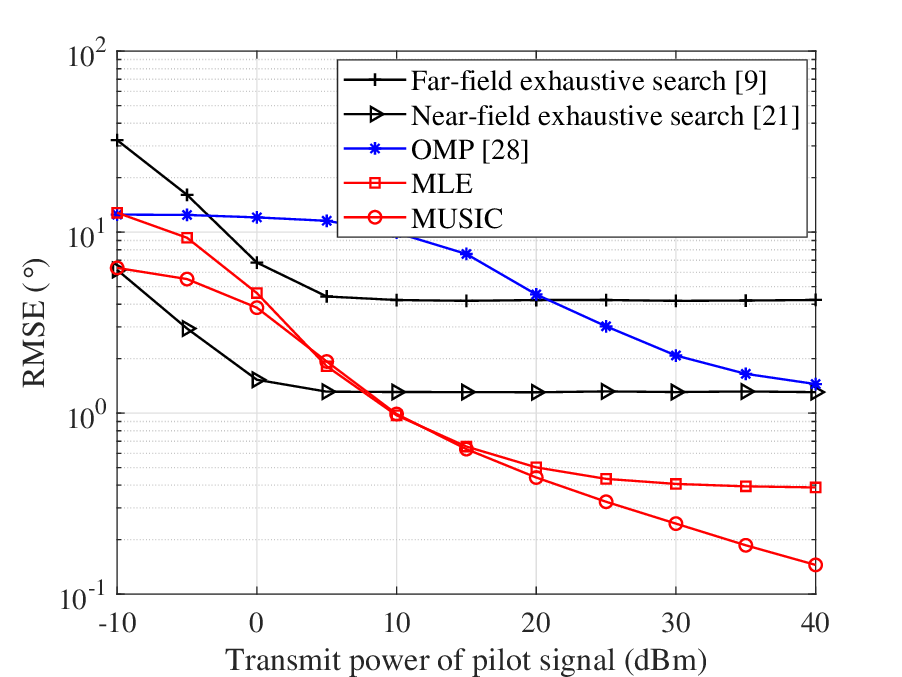}}
    \caption{Angle estimation accuracy comparison of different algorithms.}
    \label{fig_AngleErr_Comp}
\end{figure}

In Fig.~\ref{fig_AngleErr_Comp}, we compare the RMSE performance of the proposed MUSIC and MLE method with two classical direct searching-based algorithms, i.e., the far-field exhaustive search~\cite{ref_BM_3GPP} and near-field exhaustive search~\cite{ref_codebookdesign_Dai}.
Additionally, we include a CS-based near-field beam estimation utilizing an orthogonal matching pursuit algorithm (OMP)~\cite{ref_NF_sparse_CE}.
Given the computational complexity and memory resource constraints associated with near-field exhaustive search and near-field OMP, we set the number of antennas at the UE and within a subarray at the BS as 16. 
Similarly, we restrict the number of subcarriers to 16. For the near-field exhaustive search and OMP algorithms, the distance domain grid count is set at $C_d=100$.
For beam estimation algorithms, we perform the subarray-wise exhaustive search beam training to obtain the channel observation. 

As illustrated in Fig.~\ref{fig_AngleErr_Comp}, in contrast with all benchmark algorithms, the proposed MUSIC and MLE methods perform the best and achieve the lowest RMSE among the listed algorithms when the transmit power of the pilot signal exceeds 10~dBm. 
Specifically, at 15~dBm transmit power, the RMSE values of the MUSIC and MLE methods are $6.87^\circ$, $3.54^\circ$ and $0.67^\circ$ lower than those of the far-field exhaustive search, near-field exhaustive search, and OMP algorithms, respectively.
Although the near-field exhaustive search initially exhibits higher estimation accuracy compared to our proposed methods, its RMSE plateaus and stabilizes at 
$1.30^\circ$ as transmit power continues to increase. This phenomenon is attributed to the limited resolution inherent in the power-based direct searching schemes.
Moreover, due to the inherent mismatch between the far-field beam and the cross-field channel characteristics, the minimal angle estimation error for the far-field exhaustive search remains significantly high, registering at $4.17^\circ$.

\begin{figure}[t]
    \centering
    {\includegraphics[width= 0.4\textwidth]{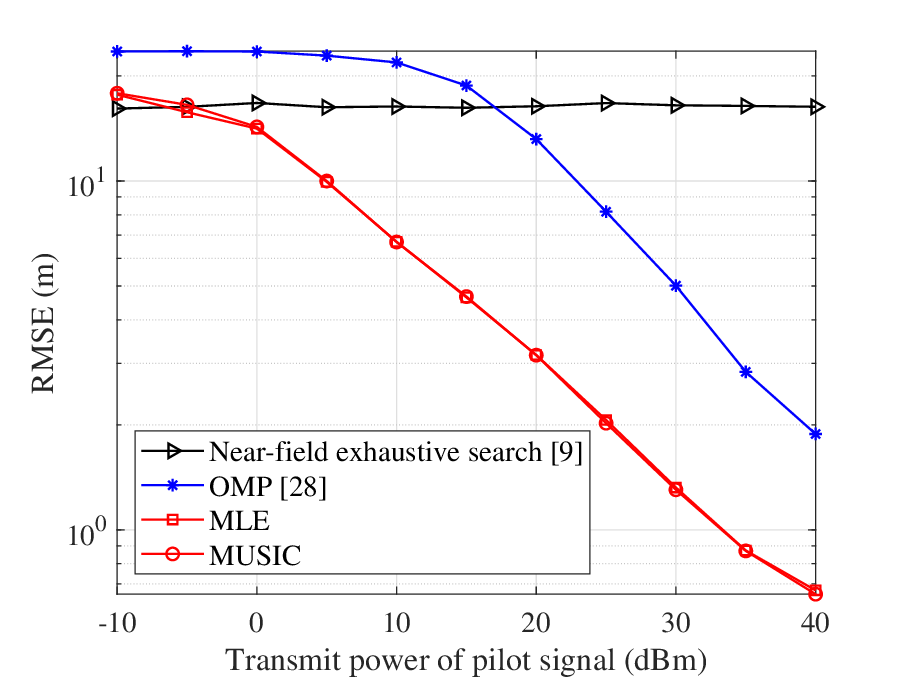}}
    \caption{Distance estimation accuracy comparison of different algorithms.}
    \label{fig_DistErr_Comp}
\end{figure}

Considering the same simulation setup, we further compare the distance estimation algorithms, as depicted in Fig.~\ref{fig_DistErr_Comp}.
Notably, the far-field exhaustive search methodology is inherently incapable of providing distance estimates.
The figure demonstrates that the distance estimation accuracy of our proposed MLE and MUSIC algorithms surpasses that of all benchmark algorithms. Specifically, the RMSE values for these methods reach as low as 0.63~m at a transmit power of 40~dBm. 
In contrast, the corresponding RMSE values for the near-field exhaustive search and OMP algorithms stand at 16.32 and 1.88~m at the same power level, respectively. 

\subsection{Beam Alignment Performance}

\begin{figure}[t]
    \centering
    {\includegraphics[width= 0.4\textwidth]{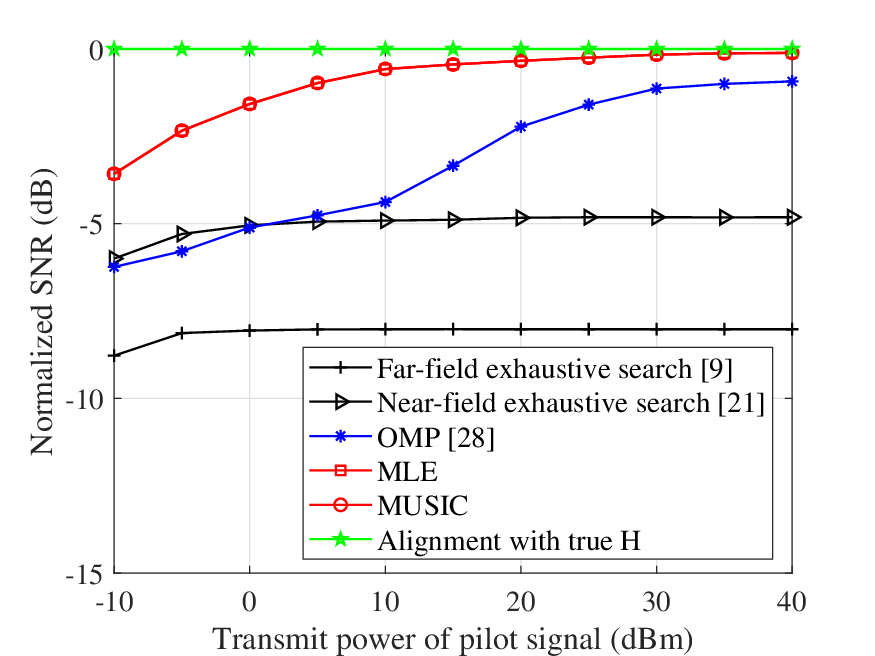}}
    \caption{SNR comparison after alignment for different methods.}
    \label{fig_SNRComp}
\end{figure}

The beam alignment problem in~\eqref{equ_training_opt} aims at finding the optimal beams to maximize the received power or SNR.
Therefore, we further evaluate the SNR after alignment. 
In Fig.~\ref{fig_SNRComp}, we evaluate the normalized SNRs achieved through angle and distance alignment based on various beam alignment schemes. 
Here, the normalized SNR is defined as the SNR after alignment over the SNR by aligning to the perfect direction decided by the channel matrix.
This ideal SNR serves as an upper bound for the performance evaluation.
In the far-field exhaustive search, the beam is directly aligned to the estimated angle.
Compared to the listed methods, the proposed alignment with MLE and MUSIC algorithms attain near-optimal beam alignment when the transmit power for the pilot signal surpasses 20~dBm. Specifically, the alignment SNR is merely 0.11~dB lower than that of the upper bound. 
This is attributable to the high-precision beam estimation executed in both the angle and distance domains.
In contrast, the SNRs resulting from alignment based on the OMP, near-field exhaustive search, and far-field exhaustive search algorithms are significantly lower than the upper bound, with values of 0.92~dB, 4.81~dB and 8.01~dB, respectively. 
This is due to the angle or distance domain beam misalignment caused by estimation errors, as illustrated in Fig.~\ref{fig_AngleErr_Comp} and Fig.~\ref{fig_DistErr_Comp}, respectively.

\section{Conclusion}
\label{sec_conclusion}
In this paper, we addressed the problem of cross-field beam alignment in THz UM-MIMO systems consisting of beam training and beam estimation steps. 
In the beam training, we have shown that it is feasible to deploy far-field beams for training, which enables adequate SNR in both far- and near-field scenarios for control signal reception. We stress that it is the beam estimation method that must take near-field characteristics into account, not the training signaling which merely must span all channel dimensions.
We have proposed the SCH beam training method, which significantly reduces training overhead by employing a 2D hierarchical codebook and spatial correlations among subarrays.
In terms of beam estimation, we have proposed the TPBE, which decouples angle and distance estimations to reduce computational complexity while conducting signal processing for multi-carrier signals to achieve high precision. 
Moreover, we employed MUSIC and MLE algorithms for high-precision angle estimation.
The MUSIC-based algorithm has higher accuracy at higher transmit power, while MLE achieves better computational efficiency.
The distance estimation is obtained based on phase differences among subarrays. 
Performance evaluations reveal that our proposed methods achieve near-optimal alignment performance with low training overhead and computational complexity.
In particular, the training overhead incurred by the SCH amounts to a mere 0.2\% when compared to that of the near-field exhaustive search. 
The TPBE yields RMSE values of $0.01^\circ$ and $0.02$~m for angle and distance estimations, respectively.
In addition, the SNR after beam alignment based on the estimation result is near-optimal, deviating by only $0.11$~dB from the ideal alignment.

\bibliographystyle{IEEEtran}
\bibliography{main} 

\begin{thebibliography}{10}
\providecommand{\url}[1]{#1}
\csname url@samestyle\endcsname
\providecommand{\newblock}{\relax}
\providecommand{\bibinfo}[2]{#2}
\providecommand{\BIBentrySTDinterwordspacing}{\spaceskip=0pt\relax}
\providecommand{\BIBentryALTinterwordstretchfactor}{4}
\providecommand{\BIBentryALTinterwordspacing}{\spaceskip=\fontdimen2\font plus
\BIBentryALTinterwordstretchfactor\fontdimen3\font minus
  \fontdimen4\font\relax}
\providecommand{\BIBforeignlanguage}[2]{{%
\expandafter\ifx\csname l@#1\endcsname\relax
\typeout{** WARNING: IEEEtran.bst: No hyphenation pattern has been}%
\typeout{** loaded for the language `#1'. Using the pattern for}%
\typeout{** the default language instead.}%
\else
\language=\csname l@#1\endcsname
\fi
#2}}
\providecommand{\BIBdecl}{\relax}
\BIBdecl

\bibitem{ref_TPBE_GC}
Y.~Chen, C.~Han, H.~Liu, and E.~Bj$\ddot{\text{o}}$rnson, ``{Far-Field Training
  With Estimation for Cross-Field Beam Alignment in Terahertz UM-MIMO
  Systems},'' in \emph{Proc. of IEEE Global Commun. Conf.}, Kuala Lumpur,
  Malaysia, 2023.

\bibitem{ref_TSR_THz}
T.~S. Rappaport, Y.~Xing, O.~Kanhere, S.~Ju, A.~Madanayake, S.~Mandal,
  A.~Alkhateeb, and G.~C. Trichopoulos, ``{Wireless Communications and
  Applications Above 100 GHz: Opportunities and Challenges for 6G and
  Beyond},'' \emph{IEEE Access}, vol.~7, pp. 78\,729--78\,757, 2019.

\bibitem{ref_6G_Mag}
W.~Tong and P.~Zhu, ``{6G: The Next Horizon: From Connected People and Things
  to Connected Intelligence},'' \emph{IEEE Wireless Commun.}, vol.~28, no.~5,
  pp. 8--8, Oct. 2021.

\bibitem{ref_hybrid_beamforming}
C.~Han, L.~Yan, and J.~Yuan, ``{Hybrid Beamforming for Terahertz Wireless
  Communications: Challenges, Architectures, and Open Problems},'' \emph{IEEE
  Wireless Commun.}, vol.~28, no.~4, pp. 198--204, Aug. 2021.

\bibitem{ref_THz_Old_revisit}
I.~F. Akyildiz, C.~Han, Z.~Hu, S.~Nie, and J.~M. Jornet, ``{Terahertz Band
  Communication: An Old Problem Revisited and Research Directions for the Next
  Decade},'' \emph{IEEE Trans. Commun.}, vol.~70, no.~6, pp. 4250--4285, May
  2022.

\bibitem{ref_cross}
C.~Han, Y.~Chen, L.~Yan, Z.~Chen, and L.~Dai, ``{Cross Far- and Near-field
  Wireless Communications in Terahertz Ultra-large Antenna Array Systems},''
  \emph{IEEE Wireless Commun.}, to appear 2023.

\bibitem{ref_hybrid_THz_CE}
W.~Yu, Y.~Shen, H.~He, X.~Yu, J.~Zhang, and K.~B. Letaief, ``{Hybrid Far- and
  Near-Field Channel Estimation for THz Ultra-Massive MIMO via Fixed Point
  Networks},'' in \emph{Proc. of IEEE Global Commun. Conf.}, Rio de Janeiro,
  Brazil, 2022, pp. 5384--5389.

\bibitem{ref_6G_near_focusing}
H.~Zhang, N.~Shlezinger, F.~Guidi, D.~Dardari, and Y.~C. Eldar, ``{6G Wireless
  Communications: From Far-Field Beam Steering to Near-Field Beam Focusing},''
  \emph{IEEE Commun. Mag.}, vol.~61, no.~4, pp. 72--77, April 2023.

\bibitem{ref_BM_3GPP}
M.~Giordani, M.~Polese, A.~Roy, D.~Castor, and M.~Zorzi, ``{A Tutorial on Beam
  Management for 3GPP NR at mmWave Frequencies},'' \emph{IEEE Commun. Surv.
  Tut.}, vol.~21, no.~1, pp. 173--196, 2019.

\bibitem{ref_AoSA_training}
C.~Lin, G.~Y. Li, and L.~Wang, ``{Subarray-Based Coordinated Beamforming
  Training for mmWave and Sub-THz Communications},'' \emph{IEEE J. Sel. Areas
  Commun.}, vol.~35, no.~9, pp. 2115--2126, Sept. 2017.

\bibitem{ref_Milli_TWC}
Y.~Chen, L.~Yan, C.~Han, and M.~Tao, ``{Millidegree-Level Direction-of-Arrival
  Estimation and Tracking for Terahertz Ultra-Massive MIMO Systems},''
  \emph{IEEE Trans. Wireless Commun.}, vol.~21, no.~2, pp. 869--883, Feb. 2022.

\bibitem{ref_nearfield_book_emil}
P.~Ramezani and E.~Bj$\ddot{\text{o}}$rnson, ``{Near-Field Beamforming and
  Multiplexing Using Extremely Large Aperture Arrays},'' to appear in the book
  Fundamentals of 6G Communications and Networking 2023.

\bibitem{ref_near_training_UCA}
Y.~Xie, B.~Ning, L.~Li, and Z.~Chen, ``{Near-Field Beam Training in THz
  Communications: The Merits of Uniform Circular Array},'' \emph{IEEE Wireless
  Commun. Lett.}, vol.~12, no.~4, pp. 575--579, April 2023.

\bibitem{ref_HBF_emil}
J.~Zhang, E.~Björnson, M.~Matthaiou, D.~W.~K. Ng, H.~Yang, and D.~J. Love,
  ``{Prospective Multiple Antenna Technologies for Beyond 5G},'' \emph{IEEE J.
  Sel. Areas Commun.}, vol.~38, no.~8, pp. 1637--1660, Aug. 2020.

\bibitem{ref_power_angle_est}
B.~Peng, K.~Guan, S.~Rey, and T.~Kürner, ``{Power-Angular Spectra Correlation
  Based Two Step Angle of Arrival Estimation for Future Indoor Terahertz
  Communications},'' \emph{IEEE Trans. Antenn. Propag.}, vol.~67, no.~11, pp.
  7097--7105, Nov. 2019.

\bibitem{ref_fast_training}
P.~Wang, J.~Fang, W.~Zhang, and H.~Li, ``{Fast Beam Training and Alignment for
  IRS-Assisted Millimeter Wave/Terahertz Systems},'' \emph{IEEE Trans. Wireless
  Commun.}, vol.~21, no.~4, pp. 2710--2724, 2022.

\bibitem{ref_BMW_SS}
Z.~Xiao, T.~He, P.~Xia, and X.-G. Xia, ``{Hierarchical Codebook Design for
  Beamforming Training in Millimeter-Wave Communication},'' \emph{IEEE Tran.
  Wireless Commun.}, vol.~15, no.~5, pp. 3380--3392, May 2016.

\bibitem{ref_QUPA}
B.~Ning, Z.~Chen, Z.~Tian, C.~Han, and S.~Li, ``{A Unified 3D Beam Training and
  Tracking Procedure for Terahertz Communication},'' \emph{IEEE Trans. Wireless
  Commun.}, vol.~21, no.~4, pp. 2445--2461, April 2022.

\bibitem{IEEE_Std_802_15_3c}
``{IEEE Standard for Information Technology-- Local and Metropolitan Area
  Networks-- Specific Requirements-- Part 15.3: Amendment 2:
  Millimeter-wave-based Alternative Physical Layer Extension},'' \emph{IEEE
  Std.}, pp. 1--200, Oct. 2009.

\bibitem{ref_Root_MUSIC_HDAPA}
F.~Shu, Y.~Qin, T.~Liu, L.~Gui, Y.~Zhang, J.~Li, and Z.~Han, ``{Low-Complexity
  and High-Resolution DOA Estimation for Hybrid Analog and Digital Massive MIMO
  Receive Array},'' \emph{IEEE Trans. Commun.}, vol.~66, no.~6, pp. 2487--2501,
  Jun. 2018.

\bibitem{ref_codebookdesign_Dai}
X.~Wei, L.~Dai, Y.~Zhao, G.~Yu, and X.~Duan, ``{Codebook design and beam
  training for extremely large-scale RIS: Far-field or near-field?}''
  \emph{China Commun.}, vol.~19, no.~6, pp. 193--204, Jun. 2022.

\bibitem{ref_eltraining_near}
S.~Xu, W.~Jintao, S.~Zhi, and S.~Jian, ``{Hierarchical Codebook-based Beam
  Training for Extremely Large-Scale Massive MIMO},'' \emph{arXiv preprint:
  2210.03345}, 2022.

\bibitem{ref_NF_Alexandropoulos}
G.~C. Alexandropoulos, V.~Jamali, R.~Schober, and H.~V. Poor, ``{Near-Field
  Hierarchical Beam Management for RIS-Enabled Millimeter Wave Multi-Antenna
  Systems},'' in \emph{Proc. of IEEE Sensor Array and Multichannel Signal
  Processing Workshop}, 2022, pp. 460--464.

\bibitem{ref_IRS_NF_codebook}
W.~Zhang and W.~Wang, ``{IRS-aided Indoor Millimeter-wave System: Near-field
  Codebook Design},'' in \emph{Proc. of IEEE Globecom Workshops (GC Wkshps)},
  Rio de Janeiro, Brazil, 2022, pp. 1489--1494.

\bibitem{ref_fast_near_training}
Y.~Zhang, X.~Wu, and C.~You, ``{Fast Near-Field Beam Training for Extremely
  Large-Scale Array},'' \emph{IEEE Wireless Commun. Lett.}, vol.~11, no.~12,
  pp. 2625--2629, Dec. 2022.

\bibitem{ref_near_paraest}
A.~M. Molaei, P.~del Hougne, V.~Fusco, and O.~Yurduseven, ``{Efficient Joint
  Estimation of DOA, Range and Reflectivity in Near-Field by Using Mixed-Order
  Statistics and a Symmetric MIMO Array},'' \emph{IEEE Trans. Veh. Tech.},
  vol.~71, no.~3, pp. 2824--2842, March 2022.

\bibitem{ref_CE_nearfield_Dai}
M.~Cui and L.~Dai, ``{Channel Estimation for Extremely Large-Scale MIMO:
  Far-Field or Near-Field?}'' \emph{IEEE Trans. Commun.}, vol.~70, no.~4, pp.
  2663--2677, April 2022.

\bibitem{ref_NF_sparse_CE}
X.~Zhang, H.~Zhang, and Y.~C. Eldar, ``{Near-Field Sparse Channel
  Representation and Estimation in 6G Wireless Communications},'' \emph{IEEE
  Trans. Commun.}, to appear 2023.

\bibitem{ref_HSPM}
Y.~Chen, L.~Yan, and C.~Han, ``{Hybrid Spherical-and Planar-Wave Modeling and
  DCNN-powered Estimation of Terahertz Ultra-massive MIMO Channels},''
  \emph{IEEE Trans. Commun.}, vol.~69, no.~10, pp. 7063--7076, Oct. 2021.

\bibitem{ref_WSMS}
L.~Yan, Y.~Chen, C.~Han, and J.~Yuan, ``{Joint Inter-Path and Intra-Path
  Multiplexing for Terahertz Widely-Spaced Multi-Subarray Hybrid Beamforming
  Systems},'' \emph{IEEE Trans.Commun.}, vol.~70, no.~2, pp. 1391--1406, Feb.
  2022.

\bibitem{ref_Fraunhofer}
K.~T. Selvan and R.~Janaswamy, ``{Fraunhofer and Fresnel Distances: Unified
  derivation for aperture antennas},'' \emph{IEEE Antennas Propag. Mag.},
  vol.~59, no.~4, pp. 12--15, 2017.

\bibitem{ref_MLE}
B.~Tang, J.~Tang, Y.~Zhang, and Z.~Zheng, ``{Maximum likelihood estimation of
  DOD and DOA for bistatic MIMO radar},'' \emph{Signal Processing}, vol.~93,
  no.~5, pp. 1349--1357, May 2013.

\bibitem{ref_PHY_526G}
T.~Levanen, O.~Tervo, K.~Pajukoski, M.~Renfors, and M.~Valkama, ``{Mobile
  Communications Beyond 52.6 GHz: Waveforms, Numerology, and Phase Noise
  Challenge},'' \emph{IEEE Wireless Commun.}, vol.~28, no.~1, pp. 128--135,
  Feb. 2021.

\bibitem{ref_QuaDRiGa}
S.~Jaeckel, L.~Raschkowski, K.~Borner, and L.~Thiele, ``{QuaDRiGa: A 3-D
  Multi-Cell Channel Model With Time Evolution for Enabling Virtual Field
  Trials},'' \emph{IEEE Trans. Antenn. Propagation}, vol.~62, no.~6, pp.
  3242--3256, Mar. 2014.

\bibitem{ref_Atrium}
Y.~Li, Y.~Wang, Y.~Chen, Z.~Yu, and C.~Han, ``{300 GHz Channel Measurement and
  Characterization in the Atrium of a Building},'' in \emph{Proc. of European
  Conference of Antennas and Propagation}, Mar. 2023, pp. 1--6.

\end{thebibliography}
\end{document}